\documentclass{article}

\usepackage{arxiv}

\usepackage[utf8]{inputenc} 
\usepackage[T1]{fontenc}    
\usepackage{url}            
\usepackage{booktabs}       
\usepackage{amsfonts}       
\usepackage{nicefrac}       
\usepackage{microtype}      
\usepackage{lipsum}		    
\usepackage{newtxmath}
\usepackage{graphicx}
\usepackage{doi}
\usepackage{tabularx, booktabs}
\usepackage{amsmath}
\usepackage{url}
\usepackage{placeins}

\newcommand{\Figure}{Fig.}
\newcommand{\Figures}{Figs.}
\newcommand{\Table}{Tab.}
\newcommand{\Tables}{Tabs.}

\title{Demography-independent behavioural dynamics influenced the spread of COVID-19 in Denmark}

\author{
    \href{https://orcid.org/0009-0009-8897-1971}{\includegraphics[scale=0.06]{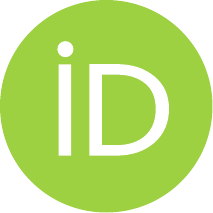}\hspace{1mm}Léo Meynent$^1$} \quad
    \href{https://orcid.org/0000-0002-6782-5635}{\includegraphics[scale=0.06]{orcid.pdf}\hspace{1mm}Michael Bang Petersen$^2$} \quad
    \href{https://orcid.org/0000-0001-6099-2345}{\includegraphics[scale=0.06]{orcid.pdf}\hspace{1mm}Sune Lehmann$^{3,4}$} \quad
    \href{https://orcid.org/0000-0001-7414-8823}{\includegraphics[scale=0.06]{orcid.pdf}\hspace{1mm}Benjamin F. Maier$^{3,4}$} \\
    $^1$AIML Lab, University of St. Gallen, Switzerland \\
    $^2$Department of Political Science, Aarhus University, Denmark \\
    $^3$DTU Compute, Technical University of Denmark \\
    $^4$Center for Social Data Science, University of Copenhagen, Denmark \\
	Correspondence: \texttt{leo.meynent@unisg.ch} \\
}

\renewcommand{\headeright}{}
\renewcommand{\undertitle}{}
\renewcommand{\shorttitle}{}

\hypersetup{
    pdftitle={Demography-independent behavioural dynamics influenced the spread of COVID-19 in Denmark},
    pdfsubject={physics.soc-ph, q-bio.PE}
    pdfauthor={Léo Meynent, Michael Bang Petersen, Sune Lehmann, Benjamin F. Maier},
    pdfkeywords={Human behaviour, COVID-19, High-frequency survey, Outbreak prediction},
}

\begin{document}
\maketitle

\begin{abstract}
	Understanding the factors that impact how a communicable disease like COVID-19 spreads is of central importance to mitigate future outbreaks.
    Traditionally, epidemic surveillance and forecasting analyses have focused on epidemiological data but recent advancements have demonstrated that monitoring behavioural changes may be equally important.
    Prior studies have shown that high-frequency survey data on social contact behaviour were able to improve predictions of epidemiological observables during the COVID-19 pandemic. Yet, the full potential of such highly granular survey data remains debated.
    Here, we utilise daily nationally representative survey data from Denmark collected during 23 months of the COVID-19 pandemic to demonstrate two central use-cases for such high-frequency survey data. First, we show that complex behavioural patterns across demographics collapse to a small number of universal key features, greatly simplifying the monitoring and analysis of adherence to outbreak-mitigation measures. Notably, the temporal evolution of the self-reported median number of face-to-face contacts follows a universal behavioural pattern across age groups, with potential to simplify analysis efforts for future outbreaks. Second, we show that these key features can be leveraged to improve deep-learning-based predictions of daily reported new infections. In particular, our models detect a strong link between aggregated self-reported social distancing and hygiene behaviours and the number of new cases in the subsequent days.
    Taken together, our results highlight the value of high-frequency surveys to improve our understanding of population behaviour in an ongoing public health crisis and its potential use for prediction of central epidemiological observables.
\end{abstract}


\section{Introduction}

Public health policy-making in response to infectious-disease outbreaks, such as COVID-19, traditionally depends on epidemiological data to assess the dynamics of disease spread and guide the implementation of non-pharmaceutical interventions (NPIs), especially in the absence of effective pharmaceutical interventions like vaccines~\cite{Flaxman2020, Brauner2021}.  
Epidemiological data, e.g.\ daily case counts, hospitalisations, and death rates come, however, with limitations when considered alone. There are reporting delays and underreporting, but more importantly, the time needed to collect and process data can cause challenges with respect to accurately predicting the spread of the disease and issuing appropriate responses in a timely manner.

Moreover, public health officials need to understand how (i) different demographics, e.g.\ age cohorts, respond and adhere to public health recommendations as well as how (ii) changes in the behaviour of these demographic groups may influence the dynamics of an infectious-disease outbreak in the short term.
Human behaviour---particularly actions like adherence to hygiene guidelines, mask-wearing, and the number of face-to-face contacts---plays a critical role in how communicable diseases like COVID-19 spread through heterogeneous populations. While first advances in outbreak progression prediction have used mobility data as proxies for population behaviour ~\cite{Benita2021, Alessandretti2022, Schlosser2022,Ruediger2021}, other studies suggested that using diary surveys asking people to report their daily number of close face-to-face encounters lead to both higher predictive power and more insights \cite{StefanScholzRStudy, Betsch2022,Koher2023}.

In particular, Koher et al. \cite{Koher2023} leveraged data from the HOPE (``How Democracies Cope With COVID-19'') project, which conducted a daily nationally representative survey in Denmark from May 2020 to March 2022 capturing granular information about the population's behaviours. They showed that self-reported number of face-to-face contacts holds more predictive power than auxiliary mobility data. At the same time, however, methodological discussions have emerged concerning both the reliability of self-reported data on behaviour due to psychological biases related to memory and social desirability \cite{hansen2022reporting, larsen2020survey} and how the behavioural dynamics of different sociodemographic groups (e.g., age groups) are reliably measured and integrated into forecasting models \cite{caselli2021mobility, monod2021age}. Such debates highlight the need for further examination and validation of the potential of survey data as a tool for epidemiological surveillance.

In this study, we utilise data from the HOPE project further, covering not only face-to-face contacts but also, for instance, adherence to hygiene practices, the use of protective measures like face masks, trust towards government decisions, and risk perception. On this basis, we present (i) new ways to analyse and categorise temporal survey data on self-reported responses during an epidemic and (ii) show how predictive models for outbreak progression can be advanced by including such data.

We first propose a new method to infer a meaningful distribution for the number of face-to-face contacts from self-reported data, which has not been achieved before. Then, we demonstrate that the temporal evolution of its median follows the same pattern across age groups. Clustering data from answers to remaining survey questions, we subsequently find four distinct categories with similar temporal behaviour across age groups, greatly reducing the number of dimensions necessary to understand how psychological responses and self-reported behaviour changed over the course of the pandemic in Denmark. We use these reduced features to predict the number of new reported cases of COVID-19 in a leave-one-out setting, showing which behavioural or auxiliary features have the most influence on the forecast by means of a state-of-the-art deep-learning model.

\section{Methods}

\textbf{Collecting representative longitudinal survey data.} We rely on the same data source as Koher et al.~\cite{Koher2023}: survey data collected on a daily basis in Denmark from May 2020 to April 2022 on a demographically representative sample of the population. The data collection was conducted within the scope of the HOPE (``How democracies cope with COVID-19'') research project, which aims to understand perceptions and behaviours of the Danish population with regard to the COVID-19 pandemic~\cite{hopeProject}. The survey was conducted in Danish in the period between July 2020 to December 2021.

Interviewees were selected through stratified random sampling based on age, sex, and geographical location, obtained directly from the Danish central person register. Each day, a representative sample of the population (restricted to residents aged 18 and older) was sent an invitation to fill out the survey via \textit{eBoks}, the official Danish electronic mail for official public communication. Participation was voluntary, and around 500 responses per day were collected, with a response rate of around 25\%. Around 8\% of the Danish population---mainly older people---were exempt from using \textit{eBoks} during the study~\cite{Jorgensen2021}.

In addition to the survey, we use publicly available epidemiological data: regional number of new cases, hospitalisations, deaths and vaccinations per day from the Danish \textit{Statens Serum Institut} (SSI)~\cite{ssiData}. We also use average national temperatures from the Danish Meteorological Institute~(DMI)~\cite{dmiData}, excluding measurements from Greenland and the Faroe Islands.

\textbf{Building a robust estimator for the median number of contacts.}
Participants were able to report the daily number of contacts with any integer, leading to a non-negligible number of outlier replies with (i) values of more than 1,000 contacts per day as well as (ii) negative data points. While the latter can be easily dismissed, there is no clear method to disregard high-valued outliers. These skew the tail of the distribution, however, and make mean and variance unreliable observables for the temporal evolution of behaviour. Alternatively, the median is a stable representation. Yet, because the domain is integer-valued and most of the replies are below 20 contacts per day with a high number of respondents reported having had exactly zero contacts on a given day, the temporal evolution of the median fluctuates between low-valued integer numbers, obfuscating subtle changes in the population's behaviour. These limitations therefore call for a more robust quantitative representation of the empirical distribution.

First, we recognise that zero-contact responses are valid, yet behaviourally distinct from non-zero responses: There is a binary latent decision between participating in contact-comprising activities or not. Second, while responses were integer values by construction, the underlying substrate for communicable disease to spread is not a natural number: transmission probability typically increases with both decreased distance and increased duration of a contact. Therefore, the integer responses in the survey have to be regarded as nearest-integer approximations of continuous variables. 

Hence, we introduce the following procedure to estimate a stable representation of the distribution of responses. To have a sufficient amount of data and to reduce the influence of weekly fluctuations, we collect the responses of the daily number of contacts for a weekly period $[t-3,t+3]$ around day $t$, for any given demographic group $a$. Then, we assume a mixed distribution consisting of (i) a point mass at zero with probability $p_{a,t}$ and (ii) a continuous log-normal distribution for positive contact numbers. For a discussion on the choice of the underlying distribution and illustration of weekly empirical distributions, see SI, Sec.~2 and \Figures{}~S10--S17. The corresponding probability density function of reported contact numbers in this period can be written as
\begin{align}
    g_{a,t}(x) = p_{a,t} \delta(x) + (1 - p_{a,t}) f_{a,t}(x),
\end{align}
where $\delta(x)$ is the Dirac delta function and $f(x)$ is the log-normal distribution. We equate $p_{a,t}$ with the proportion of non-negative responses in demographic group $a$ that indicated having had zero contacts on days lying in $[t-3,t+3]$ and find $f_{a,t}(x)$ by means of a maximum-likelihood fit of a log-normal distribution with parameters $\mu$ and $\sigma$ to the subset of strictly positive responses during this time. If $p_{a,t}\geq1/2$, the median of the full distribution is equal to zero. Otherwise, the median is given by the point where the cumulative distribution function of $f_{a,t}$ equals the remaining proportional probability $q_{a,t} = (1/2-p_{a,t})/(1-p_{a,t})$. Therefore, we have the median
\begin{align}
    \operatorname{med}[X_{a,t}] =
    \begin{cases}
        0 & p_{a,t} \geq 1/2 \\
    \exp \left(\mu +{\sqrt {2\sigma ^{2}}}\mathrm {erf} ^{-1}\left(\frac{1-2p_{a,t}}{1-p_{a,t}}-1\right)\right)  &p_{a,t} < 1/2,
    \end{cases}
\end{align}
as our estimator for the contact number on day $t$, with $\operatorname{erf}^{-1}$ being the inverse error function. As discussed in the SI, Sec.~2, this procedure reliably incorporates the tail of the distribution without being influenced by outliers of unrealistically large values.

As detailed in the Results section, a universal, demography-independent temporal contact behaviour modulation can be inferred from the age-group-specific median time series.

\textbf{Clustering survey responses into interpretable behavioural features.} We focus on a total of $34$ survey questions that were asked during the entire course of the survey. In order to regroup them into meaningful features, we use the non-negative matrix factorisation (NMF) dimensionality reduction algorithm~\cite{lee2000algorithms}, which clusters features of high similarity. 

In this dimensionality reduction, questions that are positively correlated with one another are assigned to a same category, which represents a behavioural pattern. By observing which survey questions have high or low weights in these categories, we can assign them an intrinsic meaning. We provide details on our clustering algorithm in the SI, Sec.~S1.

We find four clusters of behavioural patterns, hereinafter called \emph{features}. All of the clustered questions are of ordinal nature, with a Likert scale of values $1$ to $7$. We associate each feature with a numerical value by averaging over the responses to all questions included in the corresponding cluster. Based on the nature of the clustered questions (see SI, Sec.~1), we name the four features \textit{life degradation score}, \textit{authorities trust score}, \textit{individual compliance score}, and \textit{others' compliance score}. As we did for number of contacts, we infer a universal, demography-independent time series for each of the features (see Results section). 

\textbf{Leveraging behavioural features to predict the spread of the disease.} 
We use the demography-independent behavioural time-series to build a time-series predictor of the number of reported cases per 100k inhabitants. Given a point in time $t$ (integer representing number of days), the predictor takes as input the corresponding data of days  $t'\in \left[t-6, t\right]$ and predicts the number of daily new cases per 100k inhabitants for times $t+1$ to $t+14$. In addition to cases, we include the following input features: estimated number of contacts, \textit{life degradation score}, \textit{authorities trust score}, \textit{individual compliance score}, \textit{others' compliance score}, average temperature in Denmark, proportion of fully vaccinated individuals in Denmark, hospitalisations per 100k inhabitants, and deaths per 100k inhabitants. Denmark is divided into five administrative regions, for every trial we use one as a test set and the remaining four as the model's training set. All inputs are z-scored before being fed to the models, but targets are not. All time series but the ``proportion of vaccinated'' and estimated number of contacts are smoothed by means of a rolling average with a centred sliding window of seven days to reduce noise and the influence of weekly-modulated behaviour (the time series for contacts was computed using a 7-day sliding window already, see above).

\newpage

We train a long short-term memory (LSTM) model~\cite{hochreiter1997long} with a hidden size of 32, using 50 epochs, a learning rate of $10^{-2}$, and an $L_2$ regularisation term of $10^{-4}$.  Since the influence of each individual feature is not directly interpretable, we use explainable-AI algorithms for deep-learning models. Here, we rely on layer-wise relevance propagation (LRP)~\cite{Montavon2019}, an algorithm that maps the model's predicted values to the original input features, assigning them an individual relevance score by means of reverse propagation of relevance through the neural network. Positive relevance scores indicate an influence towards higher values of the outcome, and conversely for negative relevance scores. We use specific LRP adapted to recurrent neural networks~\cite{arras2017} and LSTMs in particular~\cite{warnecke2020}. We train multiple models with region Midtjylland (Central Denmark) as a target, using 200 different seeds, and compute the relevance scores linked with the predictions at $t+14$. We average the relevance scores over all samples in the test set. After this operation, we get a per-model-run average relevance score for every input feature, for every time point in our 7-days input window. Because the relevance score distributions have a temporal dependence that is similar across input features (cf SI, \Figure{}~S22), we divide each input feature's relevance scores on day $t$ in the past by the corresponding median value of the case-number feature's relevance score for day $t$, thereby normalising the distribution and decreasing its time-dependence (cf SI, \Figure{}~S23). Thus, relevance scores represent the prediction-influence in units of median case-number strength. In \Table{}~\ref{tab:relevance}, we show statistics of the joined distribution (over the 7 days of our input window).

\section{Results}

\begin{figure}[p]
\centering
\includegraphics[width=0.9\textwidth]{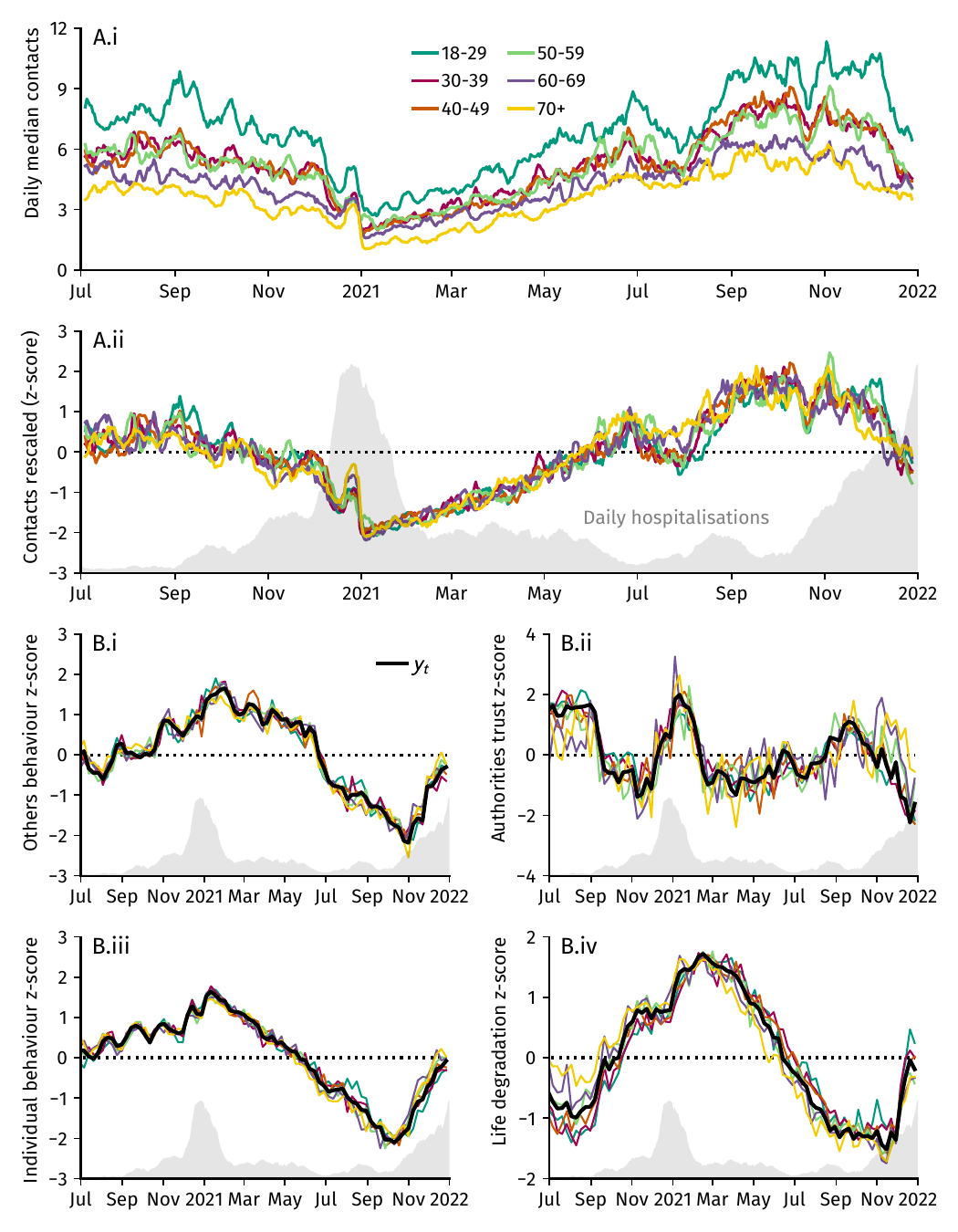}
\caption[Evolution of the daily estimated number of contacts, by age group and z-scored.]{Evolution of survey features by age groups and collapsed to the universal trend via z-scoring.  Comparison to the 7-day rolling average of the number of new daily hospitalizations (grey surface, arbitrary scale). A: Daily number of contacts by age group. \textbf{A.i}: Estimated median. \textbf{A.ii}: Median scaled by mean and standard deviation of the time series (z-scored). B: Z-scored weekly temporal evolution of remaining survey feature clusters and universal trend $y_t$. \textbf{B.i}: other's compliance. \textbf{B.ii}: Authorities trust. \textbf{B.iii}: individual compliance, \textbf{B.iv}: Life degradation.}
\label{fig:n_contacts}
\end{figure}

\textbf{Behaviours follow demography-independent dynamics across time.} We analyse the estimated number of contacts $x_{a,t}$ for each age group $a$ and day $t$.
While the various time series differ in terms of time-independent baseline and strength of variation, we also observe that the pure temporal dynamics follow a similar pattern, cf.~\Figure{}~\ref{fig:n_contacts}, panel A. When normalising the respective time series by temporal average $\mu_a=(1/T)\sum_{t=1}^T x_{a,t}$ and corresponding standard deviation $\sigma_a$, a universal temporal evolution of contact behaviour $y_{t}$ emerges, cf.\ \Figure{}~\ref{fig:n_contacts}, panel B. Here, $y_t$ is equal to the age-group average on day $t$ of their respective time series. We can reconstruct the time-dependent age-specific median contact number as $\hat{x}_{a,t} = \mu_a + \sigma_a y_t$ and find low reconstruction errors between $x_{a,t}$ and $\hat{x}_{a,t}$ (see \Table{}~\ref{tab:n_contacts}).

\begin{table}[t]
    \centering
    \caption[Reconstruction statistics for the estimated number of contacts.]{Reconstruction statistics for the estimated number of contacts, by age group. Here, ``Mean'' refers to the temporal average over the daily median number of contacts a single individual from an age group had to individuals of any age group. ``Std. Dev.'' refers to the corresponding temporal standard deviation. Column ``RMSE'' contains the root mean squared error between the respective original time series and the reconstructed time series. Similarly, column ``MAPE'' contains the mean absolute percentage error between the original and reconstructed times series.}
    \label{tab:n_contacts}
    \begin{tabular}{lrrrr}
        \toprule
         & \textbf{Mean} & \textbf{Std. Dev.} & \textbf{RMSE} & \textbf{MAPE}\\
        \midrule
        18-29 & 6.98 & 2.07 & 0.49 & 5.27\% \\
        30-39 & 5.30 & 1.62 & 0.29 & 4.56\% \\
        40-49 & 5.27 & 1.73 & 0.33 & 5.40\% \\
        50-59 & 5.25 & 1.57 & 0.35 & 5.15\% \\
        60-69 & 4.27 & 1.23 & 0.32 & 5.69\% \\
        70+ & 3.62 & 1.23 & 0.36 & 7.63\% \\
        \bottomrule
    \end{tabular}
\end{table}

While younger participants had more contacts during the pandemic on average, they also tended to modulate their behaviour more strongly, quantified by the standard deviation $\sigma_a$. In other words, when all age groups diminished their number of contacts, the younger age groups decreased theirs comparatively more; conversely, when all age groups increased their number of contacts, younger age groups increased theirs more.

We can extend this methodology to understand other demographic features, such as self-reported gender, where we note that males had a higher estimated number of contacts at $\mu=7.38$ compared to $\mu=6.23$ for females, and modulated it slightly more ($\sigma = 1.67$ compared to $\sigma = 1.47$, respectively). Going beyond demography, similar analyses can be performed for socio-economical features, such as level of education or professional occupation; we cover those in the Supplementary Information (SI), cf.\ \Figures{}~S5--S7 and \Tables{}~S1--S3.

As described in the Methods, we cluster questions that were part of the HOPE survey into four behavioural features: the \textit{life degradation score} represents the perception that quality of life has been diminished directly by the public health crisis or through NPIs for its mitigation, the \textit{authorities trust score} represents the trust that following public authorities' directives will help curb the spread of the disease, the \textit{individual compliance score} represents how much respondents adhered to hygiene and distancing recommendations, and the \textit{other's compliance score} represents respondents' perception of how diligent \emph{other} people were at applying hygiene and distancing measures.

When we apply the same decomposition procedure we used for age to these features, we find that all of them display similar age-independent modulation behaviour over time, as shown in \Figure{}~\ref{fig:n_contacts}B. The reconstruction error is generally low, (max. 3\% for weekly resolution, see SI, \Tables{}S4--S7). Similarly to the case of the number of contacts, we can characterise these features by (i) a demography-independent temporal modulation function and (ii) temporal average and standard deviation per demographic group (reported in the SI, \Tables{}S4--S7). In particular, we find that the younger age groups tend to have lower \textit{individual compliance}, \textit{authorities trust}, and \textit{other's compliance} scores than older age groups. Contrary to the case for the number of contacts, younger age groups did not tend to modulate their individual compliances more strongly than the older age groups, but their trust in authorities varied more in time, as quantified by the respective standard deviations. Regarding the \textit{life degradation} score, its average tends to be slightly higher for both the youngest and oldest age groups as compared to their middle-aged counterparts. Notably, all age groups judged their own behaviour more positively than the behaviour of others, yet the temporal modulation of the two features follows a highly similar pattern.

\textbf{Behavioural changes can be linked with infection dynamics.}
Having found a range of key, demography-independent behavioural patterns, we now study how their temporal evolution possibly influenced the spread of COVID-19 in Denmark. To this end, we leverage deep-learning-based long short-term memory (LSTM) models inspired by previous research on the topic~\cite{Du2023}.

\begin{figure*}[p]
\centering
\includegraphics[width=\textwidth]{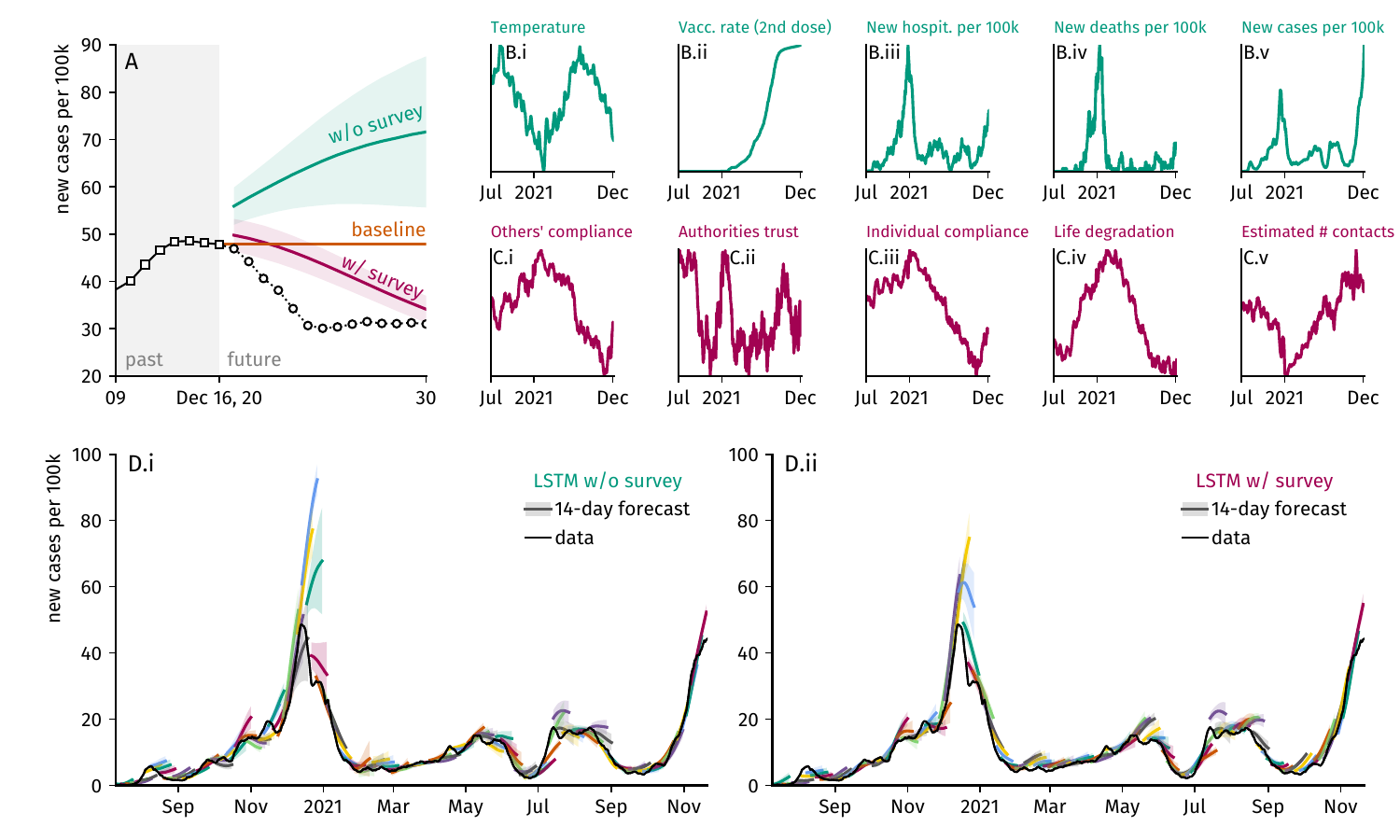}
\includegraphics[width=\textwidth]{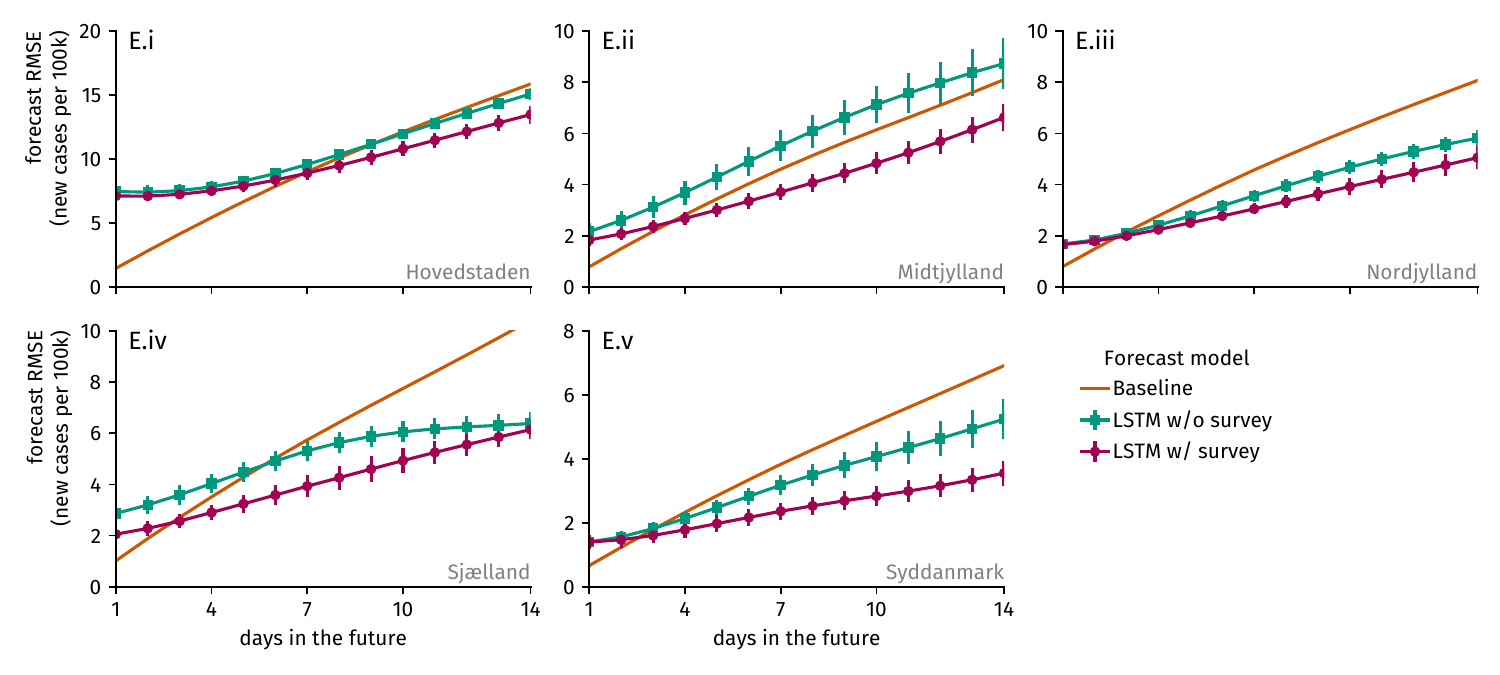}
\caption[Performance of our predictors for future reported cases, compared to baselines.]{Prediction performance of three forecast models. \textbf{A:} Example forecast for region Midtjylland for input data between Dec 9 and Dec 16, 2020. The baseline model assumes that the number of cases per 100,000 inhabitants will remain on the level of the last seen value. A long short-term memory (LSTM) model that relies only on non-survey input features (panels B.i--B.v) overestimates the future number of cases in the upcoming 14 days (band represents standard deviation over 20 independent models). An LSTM model that additionally includes survey features (panels C.i--C.v) leads to a more accurate prediction. \textbf{B:} Non-survey input features that are being used to forecast the upcoming number of new cases per 100,000 inhabitants. All timeseries but the average temperature are region-specific. Units are omitted because input features will be z-scored before being used in the respective models. \textbf{C:} Region-specific survey input features that are being used in the second LSTM model. These are the universal time series' $y_t$ that have been computed by the procedure displayed in \Figure{}~\ref{fig:n_contacts}. \textbf{D:} Example forecasts for region Midtjylland until the end of 2021, for the LSTM without survey features (D.i) and with survey features (D.ii). \textbf{E:} Average prediction root mean squared error per region, where lower values represent better fits.}
\label{fig:loo_prediction_rmse}
\end{figure*}

We compare three prediction models (cf.\ \Figure{}~\ref{fig:loo_prediction_rmse}A): (i) a baseline model that only relies on the last seen number of new cases on date $t$ and assumes that the number of cases is going to remain constant for the next 14 days, (ii) a reduced LSTM that only takes the basic features of temperature, vaccination rate, new hospitalisations, new deaths, and new cases as an input (cf.\ \Figure{}~\ref{fig:loo_prediction_rmse}B), and (iii) an LSTM that, in addition to these features, also receives all five of the demography-independent survey features as an input (cf. \Figure{}~\ref{fig:loo_prediction_rmse}C). Example forecasts for the region Midtjylland are shown in \Figure{}~\ref{fig:loo_prediction_rmse}D for both the reduced model (panel D.i) and the full model (which leverages both basic and survey features, panel D.ii). We find that on average, the full model outperforms the baseline after three days for regions Nordjylland, Sj\ae{}lland, Syddanmark, after four days for Midtjylland, and after seven days for Hovedstaden (cf. \Figure{}~\ref{fig:loo_prediction_rmse}E). Additionally, the full model outperforms the reduced model on all days, i.e.\ the survey features increase the predictive power. For a more detailed discussion see SI, Sec.~3.

\begin{table*}[t]
    \centering
    \caption[Aggregated relative relevance scores for each input feature in predicting the number of new cases in the subsequent days with an LSTM model.]{Relative relevance scores for each input feature in predicting the number of new cases in 14 days by means of 200 independent long short-term memory (LSTM) models. A positive relevance score indicates that large values of the respective feature increase the value of the prediction, while conversely a negative relevance score indicates that higher values of the respective input feature lead to a lower forecast. Because past case numbers have the highest relevance for predicting future case numbers we compute relevance scores as relative to the median case-number relevance score on day $t$ in the past, then join the distributions over all days in the past and report percentiles of this joined distribution here. This means, for instance, that the average national temperature with a median relative relevance score of -0.56 was about half as important as past case numbers in predicting future case numbers and the estimated number of contacts with a median relative relevance score of 0.29 was about one third as important. Features that have consistent influence on the forecast with all quartiles $Q_1$, $Q_2$ and $Q_3$ being either consistently positive or negative have been marked with bold text. Detailed illustrations of the relevance score distributions are shown in the SI, Figs.~S22--S23.}
    \label{tab:relevance}
    \begin{tabular}{rrrrrr}
        \toprule
             & \textbf{Q2.5} & \textbf{Q25} & \textbf{Median} & \textbf{Q75} & \textbf{Q97.5} \\
        \midrule
            \textbf{Cases / 100k} & -0.61 & 0.59 & 1.00 & 1.54 & 3.78\\
            \textbf{Average national temperature} & -3.81 & -1.02 & -0.56 & -0.24 & 0.34\\
            \textbf{Other's compliance} & -2.96 & -0.72 & -0.39 & -0.20 & 0.18\\
            \textbf{Estimated no. of contacts} & -0.26 & 0.12 & 0.29 & 0.59 & 2.11\\
            \textbf{Individual compliance} & -2.31 & -0.53 & -0.24 & -0.07 & 0.42\\
            Authorities trust & -1.16 & -0.21 & -0.01 & 0.20 & 1.57\\
            Life degradation & -0.64 & -0.10 & 0.04 & 0.16 & 0.77\\
            Hospitalisations / 100k & -0.05 & 0.00 & 0.02 & 0.05 & 0.24\\
            Deaths / 100k & -0.01 & -0.00 & 0.00 & 0.01 & 0.04\\
            Proportion of vaccinated & -0.03 & -0.01 & 0.00 & 0.01 & 0.04\\
         \bottomrule
    \end{tabular}
\end{table*}

To quantify how strongly each input feature is actually leveraged by the model to obtain a forecast, we compute a relative relevance score for each of them (cf.\ \Table{}~\ref{tab:relevance}). Because the previous numbers of new cases per 100k (unsurprisingly) have the strongest influence on the forecast, we report the relevance of other features in units of the case-number relevance. The second strongest relevance for the prediction is given by the national-average temperature which determines the forecast about half as strongly as past case numbers. All of the survey features \textit{other's compliance score} ($-39\%$ median relative relevance strength), estimated number of contacts ($29\%$ median relative relevance strength), and \textit{individual compliance score} ($-24\%$ median relative relevance strength) influence the forecast consistently. Here, higher behavioural scores of both types and lower estimated contact numbers lead to a prediction of lower new cases per 100k, respectively. Although both \textit{authorities trust} and \textit{life degradation} scores may influence the prediction (as indicated by the elevated width of their respective relative relevance strength distribution), their respective median influence is close to zero, thereby eluding a consistent interpretation. The remaining features have negligible influence.

\section{Discussion}

Understanding how an infectious disease is transmitted within a population is key for implementing effective public health policies designed to curb an outbreak, so as to limit the number of severe infections, hospitalisations, and deaths. Forecasting the spread of an infectious disease like COVID-19 is a difficult task as a multitude of factors will influence the probability of transmission. Prior work has demonstrated that longitudinal, self-reported contact data collected through surveys could improve prediction of COVID-19's spread over the use of mobility data \cite{Koher2023}. Other research, however, has argued for the importance of assessing behavioural changes over distinct demographic segments as well as raised concerns regarding the reliability of self-reported behavioural measures \cite{larsen2020survey, hansen2022reporting}.\looseness-1

We propose a first improvement of prior work by deriving a robust statistical estimator of the median daily number of contacts, which takes into account the high number of responses indicating zero contacts and remains stable despite high-valued outlier responses. This estimator allows us to perform a detailed analysis of the social distancing behaviour within different demographics. We showed that while the median number of contacts and their temporal modulation varied across age groups, those modulations were driven by a single time-varying dynamic common to every demographic, indicating that behavioural changes were rather homogeneous across the population during a large part of the pandemic in Denmark.\looseness-1

Second, we use a multitude of survey questions that were not used previously, which focus around the population's perception of the pandemic and how it affected them, in particular with regard to behaviour in accordance with health and hygiene recommendations. We suggested a simple methodology to group multiple questions into interpretable behavioural features, decreasing dimensionality drastically. Analysing their longitudinal evolution indicated that demography-specific dynamics were driven by an underlying universal time series, once again highlighting a homogeneous population-level reaction to the pandemic.
By decomposing behavioural dynamics into a demography-independent time series and demography-specific but time-independent averages and standard deviations, we found a simple yet reliable lens through which the analysis of these dynamics and their use in predictive models becomes feasible. Because infectious-disease models that explore different progression scenarios in an ongoing crisis have to rely on assumptions to model demographic behaviour such as contact modulations \cite{keeling_modeling_2008,funk_socialmixr_2018}, our finding that behavioural modulation is demography-independent may simplify such efforts in the future, too.\looseness-1
 
We subsequently used these time series to draw a link between changes in behaviour and the evolution of COVID-19 in Denmark. Our results
indicate a relationship between higher average national temperatures and a lower-valued prediction of reported COVID-19 cases in the future, which is in line with previous research that has linked lower temperatures to higher case counts~\cite{Ledebur2022}. In addition, we show a positive relation between the number of reported contacts and future cases, as well as a negative relation between better individual compliance and future reported cases. As self-reported behaviour may be subject to a social desirability bias and memory biases ~\cite{Krumpal2013, larsen2020survey, hansen2022reporting}, we also note that the \textit{others' compliance score}, which represents the respondents' perception that other people diligently followed health recommendations, is similarly associated with a decreasing influence on the number of future cases. In fact, observations of other people's behaviour were the survey feature that most strongly related to future cases, potentially because such observations may be less distorted by self-serving biases. \looseness-1

These results indicate that the reduction of contacts, as well as the adoption of safer behaviours at an individual level, are linked to a lower number of reported COVID-19 cases in the future, on a regional scale. While our methodology does not allow us to assess the causal nature of that relationship, our analysis nonetheless suggests that following public health recommendations on an individual level may have led to a reduction of the spread of the disease in Denmark.\looseness-1

While our predictors for the number of future reported cases work reasonably well, they are built in a leave-one-out setting, which implies their training set contains data about future reported cases in the four other Danish regions. Therefore, the methodology we present here cannot necessarily be reliably used to forecast case numbers in future outbreaks where of course, ``complete'' time series of neighbouring regions are unknown.\looseness-1

Nevertheless, our analysis provides valuable insights into the behaviour of the Danish population during the COVID-19 pandemic and helps understanding how it affected the spread of the disease within the country. Based on our methodology, data collected by longitudinal survey studies may be used in future outbreaks to analyse universal changes in a population's behaviour, informing decision makers or contribute to improving forecasts of key epidemiological observables.\looseness-1

\section*{Code availability}

We make our code available at \href{https://doi.org/10.5281/zenodo.14988847}{doi.org/10.5281/zenodo.14988847}.

\section*{Competing interests}

The authors declare that they have no competing interests.

\section*{Ethical considerations}

The research complied with Aarhus University’s Code of Conduct and with the ethical standards set by the Danish Code of Conduct for Research Integrity. The legal documentation underlying the survey data  was approved by Aarhus University’s Technology Transfer Office. As per section 14(2) of the Act on Research Ethics Review of Health Research Projects, ``notification of questionnaire surveys [\dots] to the system of research ethics committee system is only required if the project involves human biological material.'' All participants provided informed consent.

\section*{Acknowledgements}
For assistance with collecting the survey data, we thank Frederik Jørgensen, Alexander Bor, Marie Fly Lindholt and Louise Halberg Nielsen.

All authors are thankful to the Carlsberg Foundation who funded the study (Grant CF20-0044, HOPE: How Democracies Cope with Covid-19). Léo Meynent was supported by the Swiss Confederation through the Swiss-European Mobility Programme SEMP.


\bibliographystyle{ieeetr}
\bibliography{references} 

\begin{thebibliography}{10}

\bibitem{Flaxman2020}
S.~Flaxman, S.~Mishra, A.~Gandy, H.~J.~T. Unwin, T.~A. Mellan, H.~Coupland, {\em et~al.}, ``Estimating the effects of non-pharmaceutical interventions on covid-19 in europe,'' {\em Nature}, vol.~584, pp.~257--261, Aug 2020.

\bibitem{Brauner2021}
J.~M. Brauner, S.~Mindermann, M.~Sharma, D.~Johnston, J.~Salvatier, T.~Gavenčiak, A.~B. Stephenson, G.~Leech, G.~Altman, V.~Mikulik, A.~J. Norman, J.~T. Monrad, T.~Besiroglu, H.~Ge, M.~A. Hartwick, Y.~W. Teh, L.~Chindelevitch, Y.~Gal, and J.~Kulveit, ``Inferring the effectiveness of government interventions against covid-19,'' {\em Science}, vol.~371, no.~6531, p.~eabd9338, 2021.

\bibitem{Benita2021}
F.~Benita, ``Human mobility behavior in {COVID}-19: A systematic literature review and bibliometric analysis,'' {\em Sustainable Cities and Society}, vol.~70, p.~102916, 2021.

\bibitem{Alessandretti2022}
L.~Alessandretti, ``What human mobility data tell us about covid-19 spread,'' {\em Nature Reviews Physics}, vol.~4, pp.~12--13, Jan 2022.

\bibitem{Schlosser2022}
F.~Schlosser, B.~F. Maier, O.~Jack, D.~Hinrichs, A.~Zachariae, and D.~Brockmann, ``Covid-19 lockdown induces disease-mitigating structural changes in mobility networks,'' {\em Proceedings of the National Academy of Sciences}, vol.~117, no.~52, pp.~32883--32890, 2020.

\bibitem{Ruediger2021}
S.~Rüdiger, S.~Konigorski, A.~Rakowski, J.~A. Edelman, D.~Zernick, A.~Thieme, and C.~Lippert, ``Predicting the {SARS}-{C}o{V}-2 effective reproduction number using bulk contact data from mobile phones,'' {\em Proceedings of the National Academy of Sciences}, vol.~118, no.~31, p.~e2026731118, 2021.

\bibitem{StefanScholzRStudy}
D.~V. Tomori, N.~Rübsamen, T.~Berger, S.~Scholz, J.~Walde, I.~Wittenberg, B.~Lange, A.~Kuhlmann, J.~Horn, R.~Mikolajczyk, V.~K. Jaeger, and A.~Karch, ``Individual social contact data and population mobility data as early markers of {SARS}-{CoV}-2 transmission dynamics during the first wave in {Germany}--an analysis based on the {COVIMOD} study,'' {\em BMC Medicine}, vol.~19, p.~271, Dec. 2021.

\bibitem{Betsch2022}
V.~K. Jirsa, S.~Petkoski, H.~Wang, M.~Woodman, J.~Fousek, C.~Betsch, L.~Felgendreff, R.~Bohm, L.~Lilleholt, I.~Zettler, S.~Faber, K.~Shen, and A.~R. Mcintosh, ``Integrating psychosocial variables and societal diversity in epidemic models for predicting {COVID}-19 transmission dynamics,'' {\em PLOS Digital Health}, vol.~1, p.~e0000098, Aug. 2022.

\bibitem{Koher2023}
A.~Koher, F.~J{\o}rgensen, M.~B. Petersen, and S.~Lehmann, ``Epidemic modelling of monitoring public behavior using surveys during pandemic-induced lockdowns,'' {\em Communications Medicine}, vol.~3, p.~80, June 2023.

\bibitem{hansen2022reporting}
P.~G. Hansen, E.~G. Larsen, and C.~D. Gundersen, ``Reporting on one's behavior: a survey experiment on the nonvalidity of self-reported {COVID}-19 hygiene-relevant routine behaviors,'' {\em Behavioural Public Policy}, vol.~6, no.~1, pp.~34--51, 2022.

\bibitem{larsen2020survey}
M.~Larsen, J.~Nyrup, M.~B. Petersen, {\em et~al.}, ``Do survey estimates of the public’s compliance with {COVID}-19 regulations suffer from social desirability bias?,'' {\em Journal of Behavioral Public Administration}, vol.~3, no.~2, 2020.

\bibitem{caselli2021mobility}
F.~Caselli, F.~Grigoli, D.~Sandri, and A.~Spilimbergo, ``Mobility under the {COVID}-19 pandemic: {A}symmetric effects across gender and age,'' {\em IMF Economic Review}, vol.~70, no.~1, p.~105, 2021.

\bibitem{monod2021age}
M.~Monod, A.~Blenkinsop, X.~Xi, D.~Hebert, S.~Bershan, S.~Tietze, M.~Baguelin, V.~C. Bradley, Y.~Chen, H.~Coupland, {\em et~al.}, ``Age groups that sustain resurging {COVID}-19 epidemics in the {U}nited {S}tates,'' {\em Science}, vol.~371, no.~6536, p.~eabe8372, 2021.

\bibitem{hopeProject}
{University of Aarhus}, ``The {HOPE} project.'' \url{https://hope-project.dk/\#/about}.
\newblock (Accessed 24-Dec-2023).

\bibitem{Jorgensen2021}
F.~Jørgensen, M.~F. Lindholt, A.~Bor, and M.~B. Petersen, ``{Does face mask use elicit risk-compensation? Quasi-experimental evidence from Denmark during the SARS-CoV-2 pandemic},'' {\em European Journal of Public Health}, vol.~31, pp.~1259--1265, 08 2021.

\bibitem{ssiData}
{Statens Serum Institut}, ``Overvågningsdata for {C}ovid-19 i {D}anmark.'' \url{https://covid19.ssi.dk/overvagningsdata}.
\newblock (in Danish - Accessed 24-Dec-2023).

\bibitem{dmiData}
{Danmarks Meteorologiske Institut}, ``Dmi frie data.'' \url{https://www.dmi.dk/frie-data}.
\newblock (in Danish - Accessed 24-Dec-2023).

\bibitem{lee2000algorithms}
D.~Lee and H.~S. Seung, ``Algorithms for non-negative matrix factorization,'' {\em Advances in neural information processing systems}, vol.~13, 2000.

\bibitem{hochreiter1997long}
S.~Hochreiter, ``Long short-term memory,'' {\em Neural Computation MIT-Press}, 1997.

\bibitem{Montavon2019}
G.~Montavon, A.~Binder, S.~Lapuschkin, W.~Samek, and K.-R. M{\"u}ller, {\em Layer-Wise Relevance Propagation: An Overview}, pp.~193--209.
\newblock Cham: Springer International Publishing, 2019.

\bibitem{arras2017}
L.~Arras, G.~Montavon, K.-R. M{\"u}ller, and W.~Samek, ``{Explaining Recurrent Neural Network Predictions in Sentiment Analysis},'' in {\em Proceedings of the EMNLP 2017 Workshop on Computational Approaches to Subjectivity, Sentiment and Social Media Analysis}, pp.~159--168, Association for Computational Linguistics, 2017.

\bibitem{warnecke2020}
A.~{Warnecke}, D.~{Arp}, C.~{Wressnegger}, and K.~{Rieck}, ``Evaluating explanation methods for deep learning in security,'' in {\em 2020 IEEE European Symposium on Security and Privacy (EuroS\&P)}, 2020.

\bibitem{Du2023}
H.~Du, E.~Dong, H.~S. Badr, M.~E. Petrone, N.~D. Grubaugh, and L.~M. Gardner, ``Incorporating variant frequencies data into short-term forecasting for {COVID}-19 cases and deaths in the usa: a deep learning approach,'' {\em eBioMedicine}, vol.~89, p.~104482, 2023.

\bibitem{keeling_modeling_2008}
M.~J. Keeling and P.~Rohani, {\em Modeling infectious diseases in humans and animals}.
\newblock Princeton: Princeton University Press, 2008.
\newblock OCLC: ocn163616681.

\bibitem{funk_socialmixr_2018}
S.~Funk, L.~Willem, and H.~Gruson, ``socialmixr: {Social} {Mixing} {Matrices} for {Infectious} {Disease} {Modelling},'' Jan. 2018.
\newblock Institution: Comprehensive R Archive Network Pages: 0.4.0.

\bibitem{Ledebur2022}
K.~Ledebur, M.~Kaleta, J.~Chen, S.~D. Lindner, C.~Matzhold, F.~Weidle, C.~Wittmann, K.~Habimana, L.~Kerschbaumer, S.~Stumpfl, G.~Heiler, M.~Bicher, N.~Popper, F.~Bachner, and P.~Klimek, ``Meteorological factors and non-pharmaceutical interventions explain local differences in the spread of sars-cov-2 in austria,'' {\em PLOS Computational Biology}, vol.~18, pp.~1--16, 04 2022.

\bibitem{Krumpal2013}
I.~Krumpal, ``Determinants of social desirability bias in sensitive surveys: a literature review,'' {\em Quality {\&} Quantity}, vol.~47, pp.~2025--2047, Jun 2013.

\end{thebibliography}


\begin{thebibliography}{1}

\bibitem{mcdowell_anthropometric_2009}
C.~O. M~A~McDowell, C D~Fryar, ``Anthropometric reference data for children and adults: {U}nited {S}tates, 1988–1994,'' {\em National Center for Health Statistics. Vital Health Stat}, vol.~11, 2009.

\bibitem{clauset}
A.~Clauset, C.~R. Shalizi, and M.~E.~J. Newman, ``Power-law distributions in empirical data,'' {\em SIAM Review}, vol.~51, no.~4, pp.~661--703, 2009.

\bibitem{maier_pareto}
B.~F. Maier, ``Maximum-likelihood fits of piece-wise pareto distributions with finite and non-zero core,'' {\em arXiv}, 2023.

\end{thebibliography}

\end{document}


\maketitle

\section{Cluster identification by NMF}
\label{sec:nmf-cluster}
Consider a non-negative matrix $X$ of dimension $n_{\mathrm{samples}} \times n_{\mathrm{questions}}$ that contains information about how each participant responded to a question. We construct $X$ by percentile-transforming the Likert scale integer numbers that people responded with, i.e.\ we map each data point to its percentile in the distribution of all other responses to the same question.

In the non-negative matrix factorisation (NMF) method, one finds two non-negative matrices $W$ $(n_{\mathrm{samples}} \times N)$ and $H$ $(N \times n_{\mathrm{questions}})$ so that $X \approx W \cdot H$, with $N$ being the number of dimensions we reduce to, i.e.\ the number of features or clusters. Here, $H_{fq}$ quantifies the topic strength, i.e.\ the strength with which responses to question $q$ contribute to the feature $f$. $W_{if}$ is the topic weight, i.e.\ the strength of a respondent's reply had they replied to the feature $f$ (i.e.\ a cluster of questions).

\begin{figure}
    \centering
    \includegraphics[width=0.7\linewidth]{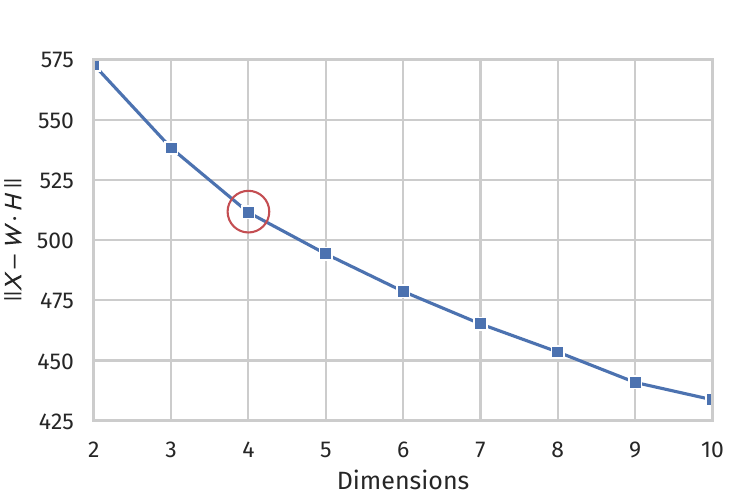}
    \caption{Frobenius norm of reconstruction error matrix $||X-W\cdot H||$ with increasing dimension $N$ (number of clusters). Marked is the elbow $N=4$.}
    \label{fig:frobenius-norm-elbow}
\end{figure}

To find $W$ and $H$ we use the NMF implementation in Python library \texttt{scikit-learn} with seed 2020, as well as parameters \texttt{max\_iter=1024}, \texttt{alpha\_W=0.002}, and \texttt{alpha\_H=0.1}. The decreasing Frobenius norm of the reconstruction error matrix $X-W\cdot H$ is shown in Fig.\ \ref{fig:frobenius-norm-elbow}. 
We chose $N=4$ by means of the elbow method with regard to this norm.

By construction, the clusters that emerge are a mixture of every question in the survey. We build behavioural features from them by keeping the highly weighted questions only: for some question $q$ and behavioural feature $f$, $f$ will include $q$ if $H_{f,q} > 0.2$. This allows a straightforward interpretation of our features based on the questions they include (see below).

\begin{figure}
    \centering
    \includegraphics[width=0.7\linewidth]{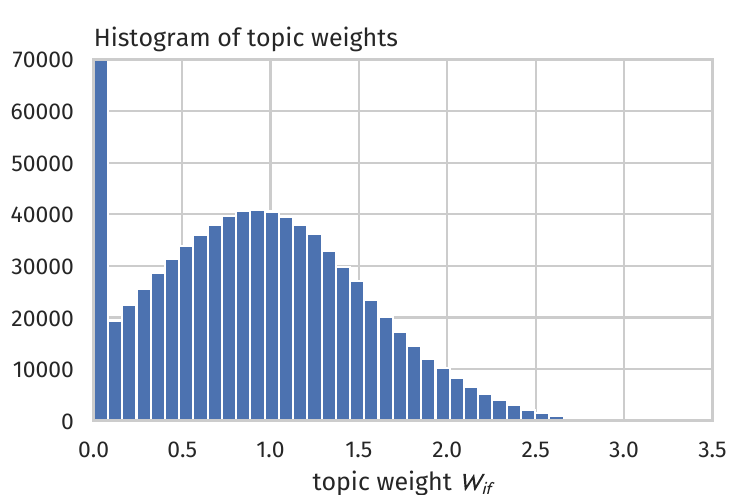}
    \caption{NMF topic weight distribution $W_{if}$.}
    \label{fig:nmf-weight-distribution}
\end{figure}

The topic weight distribution of values $W_{if}$ with $N=4$ is shown in Fig.\ \ref{fig:nmf-weight-distribution}.

\subsection{Survey questions by cluster}
\label{sec:nmf-survey-questions}

\begin{figure}
    \centering
    \includegraphics[width=\textwidth]{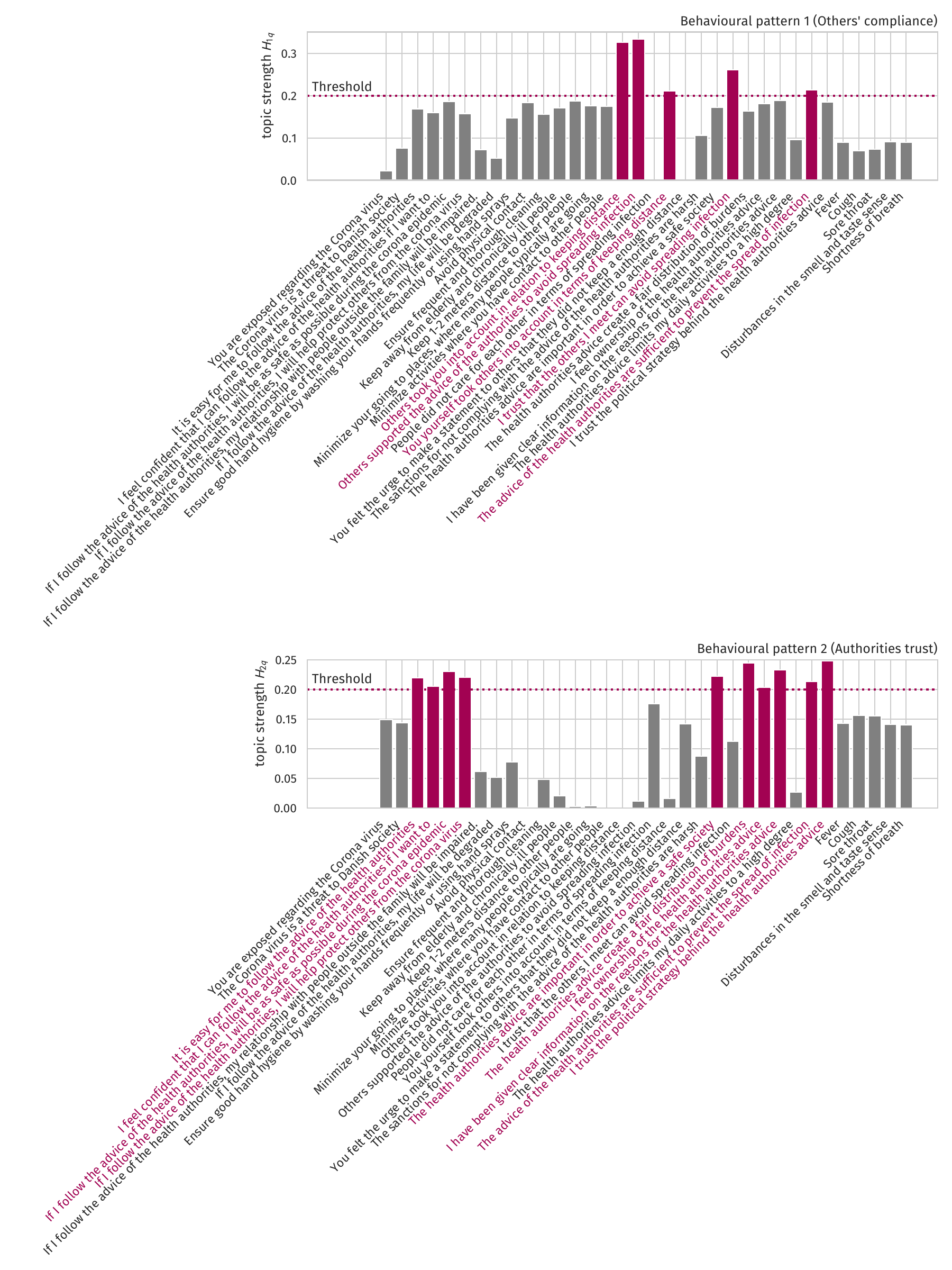}
    \caption{Topic strengths $H_{fq}$ for questions $q$ in the first two of the $N=4$ identified clusters.}
    \label{fig:question-topic-strengths-0}
\end{figure}

\begin{figure}
    \centering
    \includegraphics[width=\textwidth]{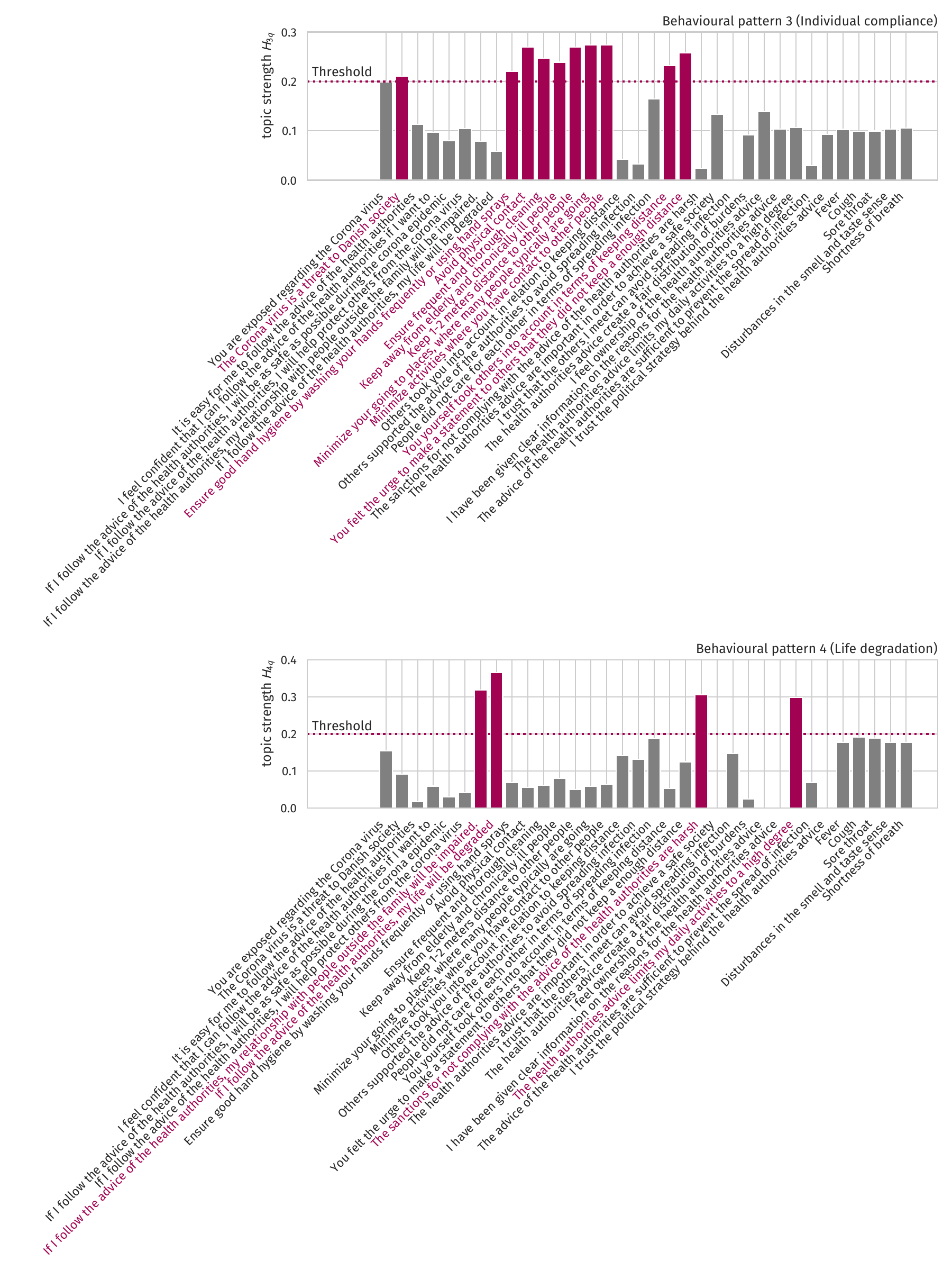}
    \caption{Topic strengths $H_{fq}$ for questions $q$ in the second two of the $N=4$ identified clusters.}
    \label{fig:question-topic-strengths-1}
\end{figure}

Question $q$ topic strengths $H_{fq}$ for the identified $N=4$ clusters are shown in Figs.\ \ref{fig:question-topic-strengths-0}--\ref{fig:question-topic-strengths-1}. With threshold $\theta_H=0.2$, the following questions are identified as the main contributors to their respective cluster.

\subparagraph{Pattern 1 (Others' compliance)}
\begin{itemize}
\item Others took you into account in relation to keeping distance
\item Others supported the advice of the authorities to avoid spreading infection
\item You yourself took others into account in terms of keeping distance
\item I trust that the others I meet can avoid spreading infection
\item The advice of the health authorities are sufficient to prevent the spread of infection
\end{itemize}

\subparagraph{Pattern 2 (Authorities trust)}
\begin{itemize}
\item It is easy for me to follow the advice of the health authorities
\item I feel confident that I can follow the advice of the health authorities if I want to
\item If I follow the advice of the health authorities, I will be as safe as possible during the corona epidemic
\item If I follow the advice of the health authorities, I will help protect others from the corona virus
\item The health authorities advice are important in order to achieve a safe society
\item The health authorities advice create a fair distribution of burdens
\item I feel ownership of the health authorities advice
\item I have been given clear information on the reasons for the health authorities advice
\item The advice of the health authorities are sufficient to prevent the spread of infection
\item I trust the political strategy behind the health authorities advice
\end{itemize}

\subparagraph{Pattern 3 (Individual compliance)}
\begin{itemize}
\item The Corona virus is a threat to Danish society
\item Ensure good hand hygiene by washing your hands frequently or using hand sprays
\item Avoid physical contact
\item Ensure frequent and thorough cleaning
\item Keep away from elderly and chronically ill people
\item Keep 1-2 meters distance to other people
\item Minimize your going to places, where many people typically are going
\item Minimize activities where you have contact to other people
\item You yourself took others into account in terms of keeping distance
\item You felt the urge to make a statement to others that they did not keep a enough distance
\end{itemize}

\subparagraph{Pattern 4 (Life degradation)}
\begin{itemize}
\item If I follow the advice of the health authorities, my relationship with people outside the family will be impaired.
\item If I follow the advice of the health authorities, my life will be degraded
\item The sanctions for not complying with the advice of the health authorities are harsh
\item The health authorities advice limits my daily activities to a high degree
\end{itemize}

\section{Fitting contact distributions}

\subsection{Estimating the maximum number of possible contacts}
The working definition for ``contact'' in the HOPE survey was an encounter that lasted at least $\tau = 15\mathrm{min}$ within a distance of $R=2\mathrm m$. If we model humans as disks of radius $r$ distributed within a circle of $R$, the number of people that fit within that circle is given as
\begin{align}
    n = \theta \frac{R^2}{r^2}
\end{align}
with $\theta\approx0.9$ being the circle packing fraction on a plane. The minimum amount of time a person has to spend within that circle to count as a ``contact'' is $\tau$. The maximum exchange rate of a person leaving the circle and a new person joining is therefore $n/\tau$. Given the maximum amount of time $T$ per day, the maximum number of close contacts per person per day is given as
\begin{align}
    \hat k = \frac{Tn}{\tau} = \frac{T\theta R^2}{\tau r^2}.
\end{align}
For the human radius, we assume $r\approx 20\mathrm{cm}$ which is approximately half the shoulder width of the average US male at the end of the 20th century \cite{mcdowell_anthropometric_2009}.

Given that there are 24h in the day, this leaves us with a maximum number of $\hat k = 8640$ per day, so at least a maximum amount of $10^4$ is a plausible assumption. We can plausibly reduce this number further by recognizing that people usually spend only about half of a day meeting other people. Further, the most common transmission pathway of diseases like COVID-19 is face-to-face, which means that only about a quarter of the circle packed with humans will count towards the number of high-risk encounters. This reduces the maximum number of possible close encounters per day to $\hat k = 1080$.

\subsection{Choosing an underlying hypothetical distribution}
As outlined in the main text, we assume a latent decision of people on whether or not to participate in contact-behavior, i.e.\ we only fit reported contact numbers $k>0$. We observe that the empirical distributions have a heavy tail (cf. Figs.~\ref{fig:weekly-contact-pdf-00}--\ref{fig:weekly-contact-pdf-07}). The canonical candidates for theoretical distributions to fit against such data are a log-normal distribution
\begin{align}
    f_L(k) = {\displaystyle \ {\frac {1}{\ k\sigma {\sqrt {2\pi \ }}\ }}\ \exp \left(-{\frac {\left(\ln k-\mu \ \right)^{2}}{2\sigma ^{2}}}\right)}
\end{align}
and a power-law (Pareto) distribution \cite{clauset}. In order to not disregard the ``core'' of the distribution, i.e.\ the part of the domain with low contact numbers $k\leq20$, we choose to compare the log-normal distribution to a piecewise Pareto distribution with an algebraic core \cite{maier_pareto}
\begin{align}
    f_P(k) = \begin{cases}
       C [2-(k/k_\mathrm{min})^\beta]& 0 \leq k \leq k_{\mathrm{min}}\\
       C (k_\mathrm{min}/k)^\alpha& k > k_\mathrm{min}.
    \end{cases}
\end{align}
We find that given the data points, the log-normal distribution generally has a higher likelihood than the Pareto distribution with algebraic core (see Figs.~\ref{fig:daily-age-group-pdfs-18-39}--\ref{fig:daily-age-group-pdfs-60+}). In cases where the Pareto distribution has a higher likelihood, the results are not significant ($p\geq0.05$ as per the procedure in \cite{clauset}).

Because the Pareto distribution fits do not necessarily find power-law tail exponents $\alpha>2$, both mean and variance of the distribution are unstable and therefore unreliable observables. However, the median of both log-normal and algebraic-core Pareto distributions generally follow the same temporal evolution (see Figs.~\ref{fig:daily-age-group-pdfs-18-39}--\ref{fig:daily-age-group-pdfs-60+}). Furthermore, once all timeseries of median, mean, and contact number $\left<k^2\right>/\left<k\right>$ are rescaled by means of z-scoring, they all follow approximately the same shape.

Looking at the results of weekly fits in Figs.~\ref{fig:weekly-contact-pdf-00}--\ref{fig:weekly-contact-pdf-07} we observe that the empirical distributions follow the tail of the log-normal distribution quite well up until values close to $k\approx\hat k\approx10^3$, at which point the reported contact number distribution deviates from the log-normal to display a stronger heavy-tail behavior. Because values $k > \hat k$ seem rather unachievable, however (as argued above), we argue that it is safe to dismiss these outliers and deem the log-normal distribution to be an appropriate representative of the empirical distributions.

\section{Properties of full distribution}
The first two moments of the full distribution as given by Eq.~(1) in the main text are
\begin{align}
    \left< x \right> &= (1-p_{a,t}) \left<x\right>_f\\
    \left< x^2 \right> &= (1-p_{a,t}) \left<x^2\right>_f,\\
\end{align}
such that variance and contact number are given by
\begin{align}
    \mathrm{Var}[x] &= (1-p_{a,t})\left[
    \left<x^2\right>_f - 
    (1-p_{a,t}) \left<x\right>_f
    \right]\\
    \frac{\left< x^2 \right>}{\left< x \right>} &=
        \frac{\left< x^2 \right>_f}{\left< x \right>_f}.  
\end{align}
Symbol $\left<\cdot\right>_f$ denotes the average over the distribution $f$ for $x>0$.
\section{Case number prediction and feature relevance}


We present results in the main text, Fig.~2. Our full-feature LSTMs outperform the LSTMs which do not use survey data in all cases, showing that the survey features are informative. Performance assessment against the naive baseline is less straightforward as it depends both on the test region and how far in the future we make the prediction. From $t+7$ onwards, our models always outperform the baseline, and in all regions but Hovedstaden (Denmark's capital region), they outperform it from $t+3$, and $t+4$ days onwards, respectively. We note an especially poor performance of our models for the Hovedstaden region: it is the most densely populated region of the country, and therefore experienced number of cases per capita higher than in the other regions. Given our models are trained on a leave-one-out basis, they failed to predict these higher number of cases as they were absent from their training set.

We note an especially poor performance of our models for the Hovedstaden region: it is the most densely populated region of the country, and the number of new cases per 100k inhabitants had more influence on the prediction than in other regions (with an average of $25$ per $100,000$ reported cases per day compared to about $12$ to $16$ in the other regions). Since our models are trained on a leave-one-out basis, they failed to predict these higher number of cases as they were absent from their training set. Notwithstanding, the predictions of our LSTM models are generally satisfying, and we focus our explanation efforts on predictions that lie further in the future where our model outperforms both comparison models.

Feature relevance scores are shown in Figs.~\ref{fig:boxplot-explanatory-strength-absolute}--\ref{fig:boxplot-explanatory-strength-relative}.

Unsurprisingly, the most relevant feature to predict future case numbers are past case numbers. We note that the time series temperature, estimated number of contacts, others and individual compliance, as well as life degradation score are highly correlated and therefore carry similar information (see Figs.~S8-S9). Nonetheless, they are leveraged with different strengths by the forecasting model.


\bibliographystyle{ieeetr}
\bibliography{references_si} 

\newpage

\FloatBarrier

\begin{figure}
    \centering
    \includegraphics[width=\textwidth]{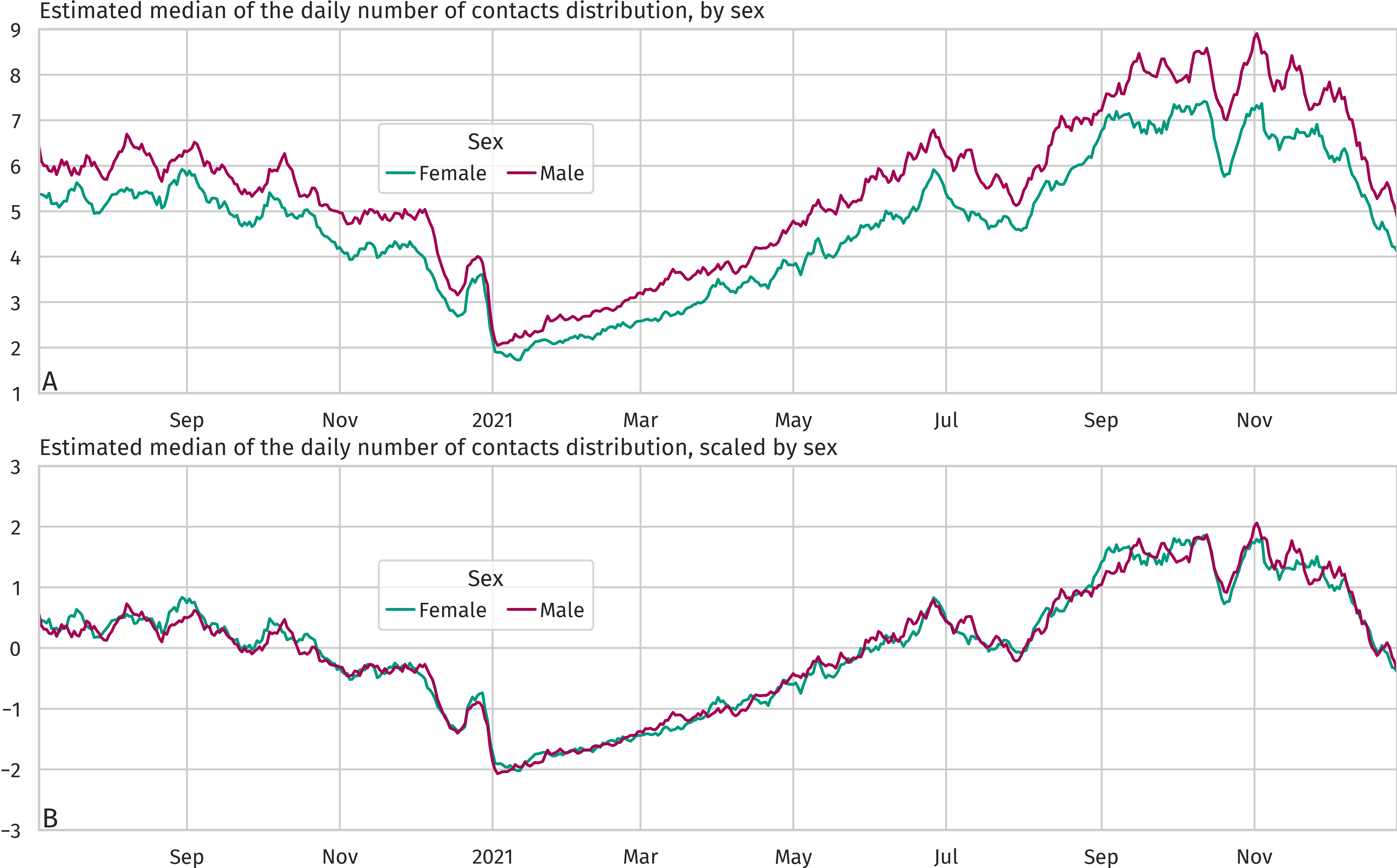}
    \caption{\textbf{Top:} Daily median contact number time series by gender (self-identified). \textbf{Bottom:} Z-scored time series. For mean, standard deviation, and reconstruction error using universal (collapsed) time series $y_t$ see Tab.\ \ref{tab:reconstruction-sex}.}
    \label{fig:contacts-sex}
\end{figure}

\begin{figure}
    \centering
    \includegraphics[width=\textwidth]{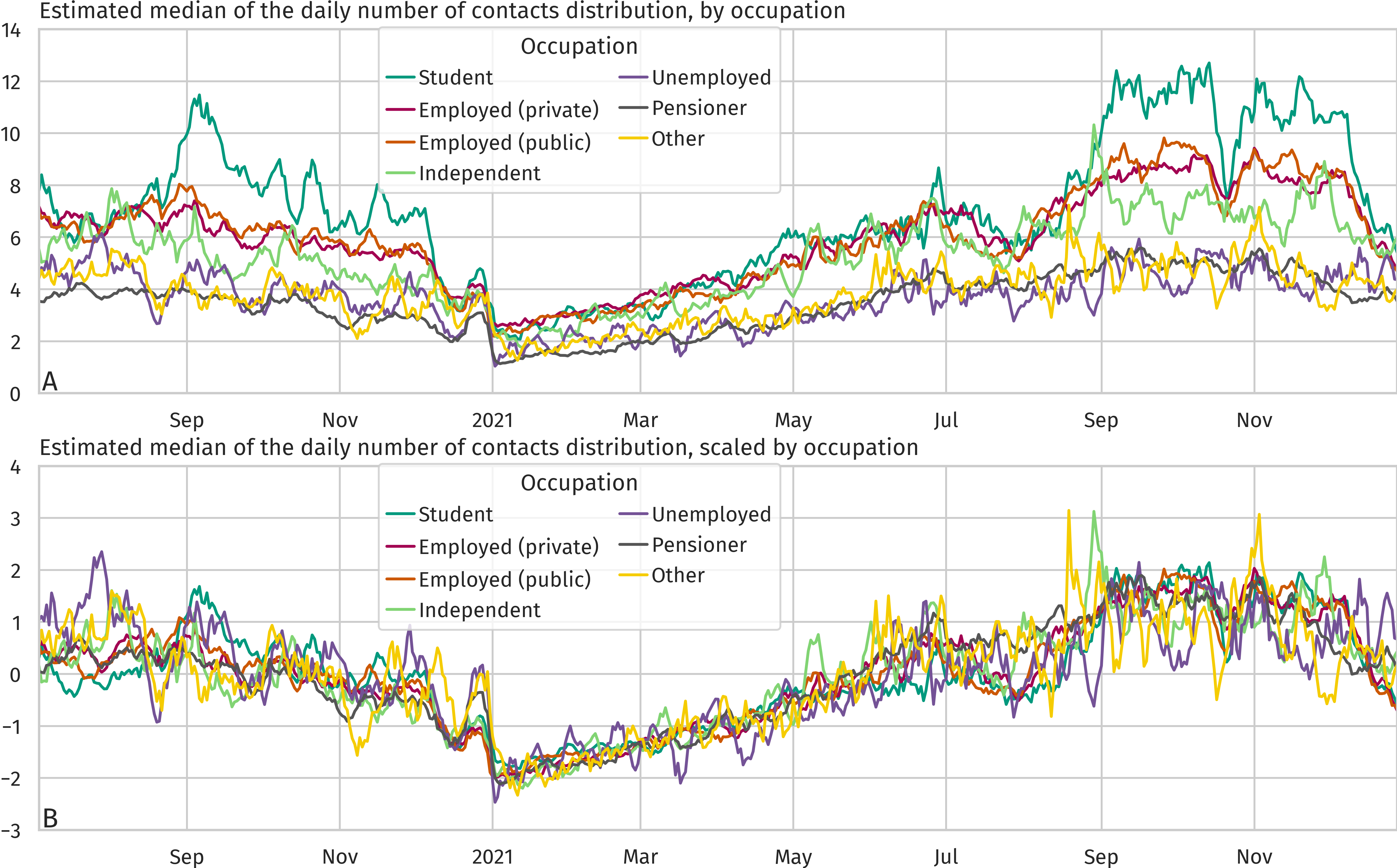}
    \caption{\textbf{Top:} Daily median contact number time series by occupation. \textbf{Bottom:} Z-scored time series. For mean, standard deviation, and reconstruction error using universal (collapsed) time series $y_t$ see Tab.\ \ref{tab:reconstruction-occupation}.}
    \label{fig:contacts-occupation}
\end{figure}

\begin{figure}
    \centering
    \includegraphics[width=\textwidth]{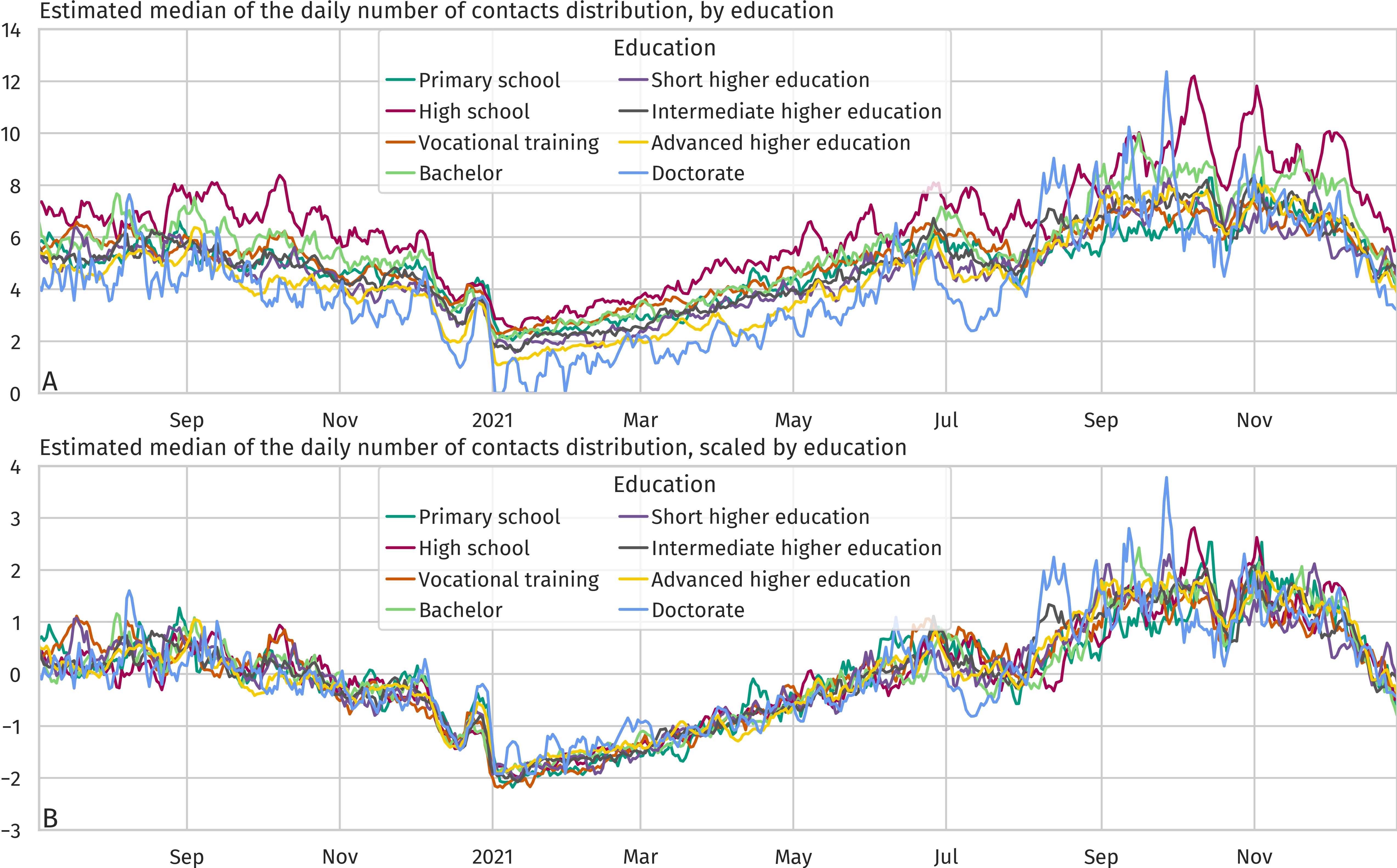}
    \caption{\textbf{Top:} Daily median contact number time series by education. \textbf{Bottom:} Z-scored time series. For mean, standard deviation, and reconstruction error using universal (collapsed) time series $y_t$ see Tab.\ \ref{tab:reconstruction-education}.}
    \label{fig:contacts-education}
\end{figure}


\begin{figure}
    \centering
    \includegraphics[width=\textwidth]{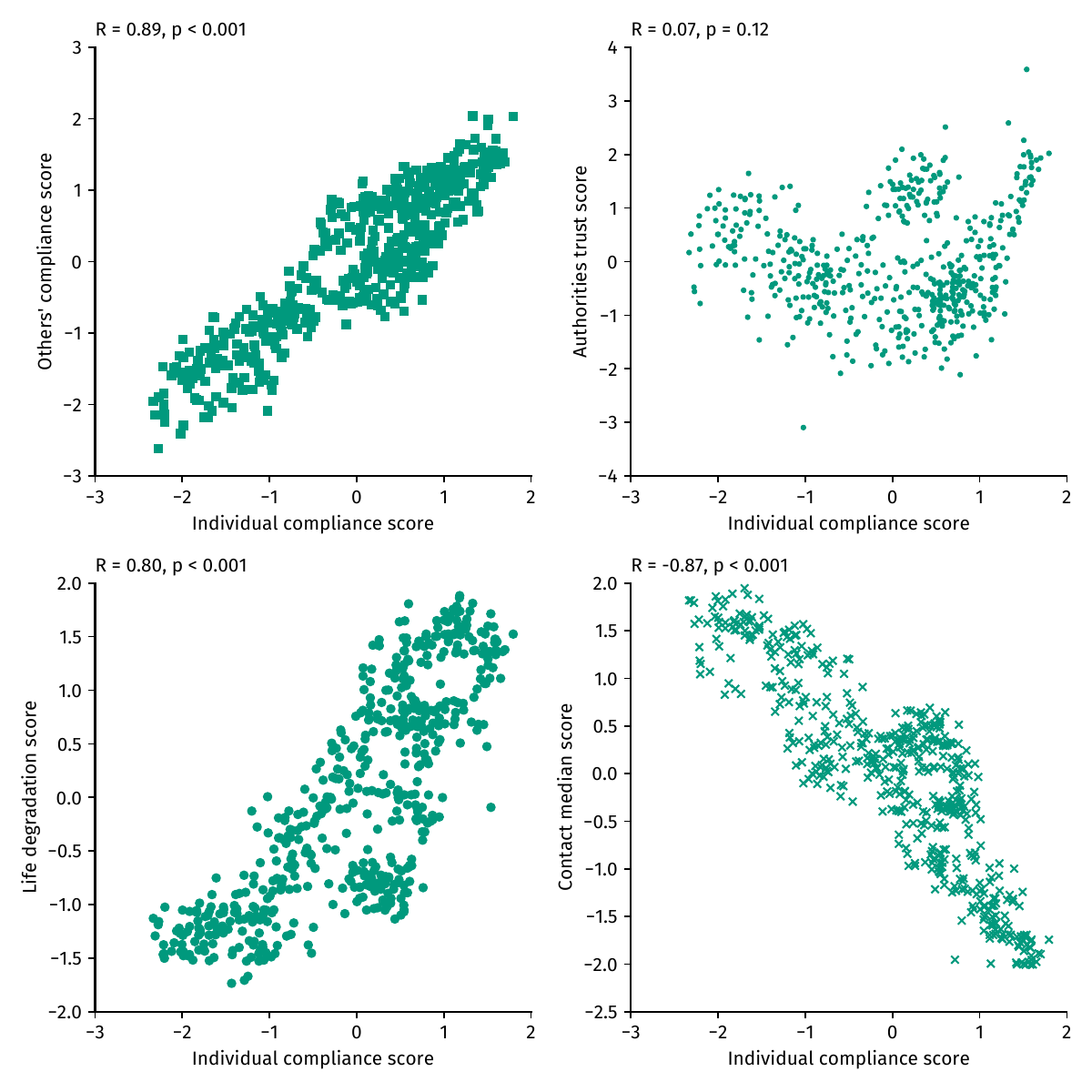}
    \caption{Pearson correlation between universal time series $y_t$ of features identified Sec.\ \ref{sec:nmf-cluster} (daily resolution).}.
    \label{fig:correlation-time-series}
\end{figure}

\begin{figure}
    \centering
    \includegraphics[width=\textwidth]{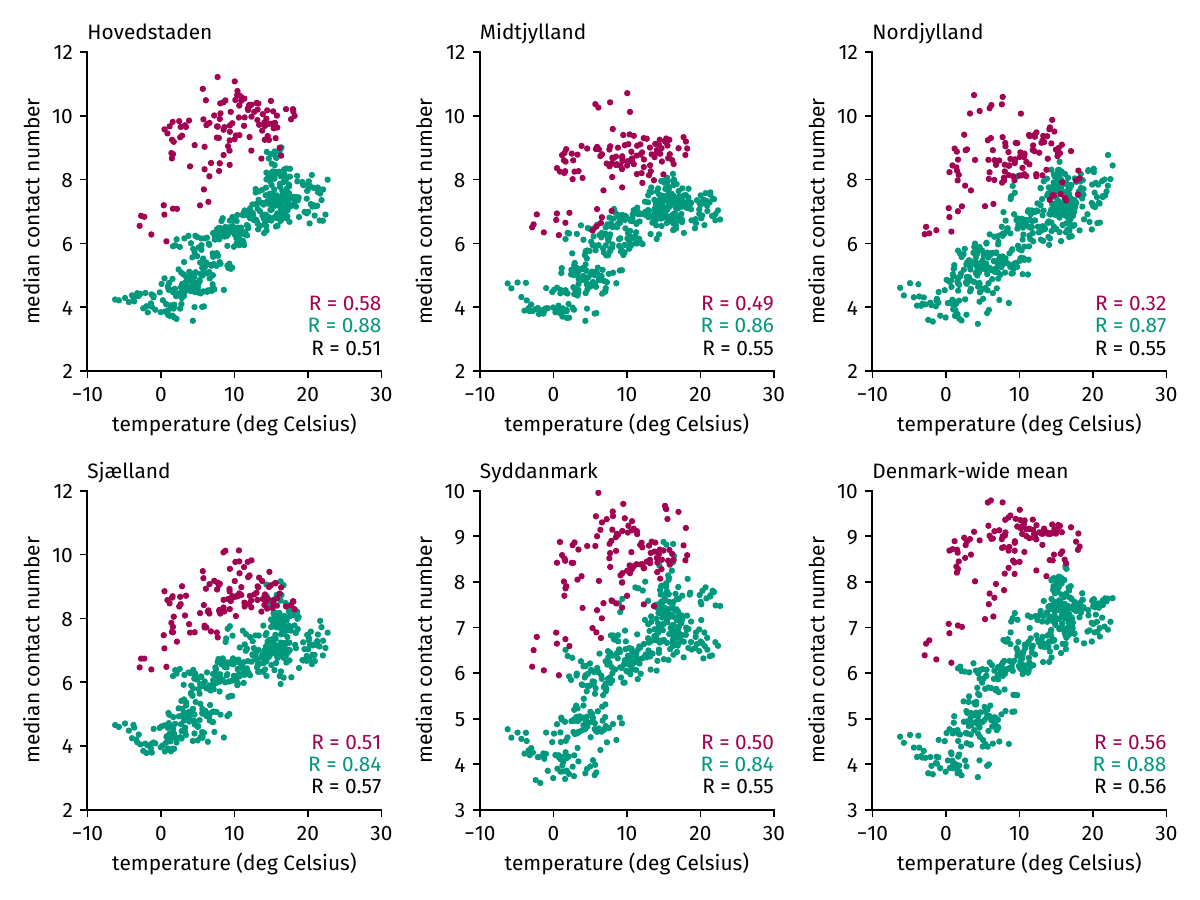}
    \caption{Correlation between universal time series $y_t$ of median contact number and temperature in each of region and Pearson correlation coefficient $R$. Teal: Before Sep 1, 2021. Magenta: After Sep 1, 2021. Black: $R$ for the entire time frame.}
    \label{fig:feature-survey-responses-mean}
\end{figure}

\begin{table}
    \centering
    \begin{tabular}{lrrrr}
        \toprule
         & \textbf{Mean} & \textbf{Std. Dev.} & \textbf{RMSE} & \textbf{MAPE}\\
        \midrule
        Female & 4.70 & 1.47 & 0.11 & 1.89\% \\
        Male & 5.49 & 1.66 & 0.14 & 1.97\% \\
        \bottomrule
    \end{tabular}
    \caption{Temporal mean and standard deviation of daily median contact number time series, by self-identified gender (cf.\ Fig.\ \ref{fig:contacts-sex} ). Also shown: Reconstruction root mean squared (RMSE) and mean average percentage error (MAPE) of the reconstruction of the respective time series based on the universal (collapsed) time series $y_t$.}
    \label{tab:reconstruction-sex}
\end{table}

\begin{table}
    \centering
    \begin{tabular}{lrrrr}
        \toprule
         & \textbf{Mean} & \textbf{Std. Dev.} & \textbf{RMSE} & \textbf{MAPE}\\
        \midrule
        Student & 6.95 & 2.69 & 0.96 & 11.67\% \\
        Employed (private) & 5.95 & 1.71 & 0.23 & 3.23\% \\
        Employed (public) & 5.97 & 1.91 & 0.39 & 5.62\% \\
        Independent & 5.28 & 1.61 & 0.65 & 8.80\% \\
        Unemployed & 3.65 & 1.06 & 0.65 & 13.87\% \\
        Pensioner & 3.48 & 1.11 & 0.30 & 6.70\% \\
        Other & 3.80 & 1.09 & 0.61 & 12.43\% \\
        \bottomrule
    \end{tabular}
    \caption{Temporal mean and standard deviation of daily median contact number time series, by occupation (cf.\ Fig.\ \ref{fig:contacts-occupation} ). Also shown: Reconstruction root mean squared (RMSE) and mean average percentage error (MAPE) of the reconstruction of the respective time series based on the universal (collapsed) time series $y_t$.}
    \label{tab:reconstruction-occupation}
\end{table}

\begin{table}
    \centering
    \begin{tabular}{lrrrr}
        \toprule
         & \textbf{Mean} & \textbf{Std. Dev.} & \textbf{RMSE} & \textbf{MAPE}\\
        \midrule
        Primary school & 4.92 & 1.33 & 0.39 & 6.20\% \\
        High school & 6.47 & 2.04 & 0.57 & 6.38\% \\
        Vocational training & 5.13 & 1.31 & 0.30 & 4.73\% \\
        Bachelor & 5.55 & 1.83 & 0.41 & 6.07\% \\
        Short higher education & 4.71 & 1.54 & 0.39 & 6.09\% \\
        Intermediate higher education & 4.94 & 1.61 & 0.28 & 4.19\% \\
        Advanced higher education & 4.50 & 1.79 & 0.32 & 6.98\% \\
        Doctorate & 4.16 & 2.17 & 0.97 & -- \\
        \bottomrule
    \end{tabular}
    \caption{Temporal mean and standard deviation of daily median contact number time series, by education (cf.\ Fig.\ \ref{fig:contacts-education} ). Also shown: Reconstruction root mean squared (RMSE) and mean average percentage error (MAPE) of the reconstruction of the respective time series based on the universal (collapsed) time series $y_t$.}
    \label{tab:reconstruction-education}
\end{table}


\begin{table}
    \centering
    \begin{tabular}{lrrrr}
        \toprule
         & \textbf{Mean} & \textbf{Std. Dev.} & \textbf{RMSE} & \textbf{MAPE}\\
        \midrule
        18-29 & 4.42 & 0.23 & 0.04 & 0.77\% \\
        30-39 & 4.59 & 0.26 & 0.04 & 0.74\% \\
        40-49 & 4.77 & 0.28 & 0.04 & 0.67\% \\
        50-59 & 4.94 & 0.27 & 0.04 & 0.65\% \\
        60-69 & 5.11 & 0.27 & 0.05 & 0.77\% \\
        70+ & 5.24 & 0.27 & 0.05 & 0.87\% \\
        \bottomrule
    \end{tabular}
    \caption{Temporal mean and standard deviation of the weekly \emph{others' compliance score} time series, by age group. Also shown: Reconstruction root mean squared (RMSE) and mean average percentage error (MAPE) of the reconstruction of the respective time series based on the universal (collapsed) time series $y_t$.}
    \label{tab:reconstruction-others-behaviour}
\end{table}

\begin{table}
    \centering
    \begin{tabular}{lrrrr}
        \toprule
         & \textbf{Mean} & \textbf{Std. Dev.} & \textbf{RMSE} & \textbf{MAPE}\\
        \midrule
        18-29 & 4.67 & 0.36 & 0.07 & 1.10\% \\
        30-39 & 4.91 & 0.37 & 0.05 & 0.77\% \\
        40-49 & 5.04 & 0.39 & 0.04 & 0.66\% \\
        50-59 & 5.31 & 0.35 & 0.05 & 0.72\% \\
        60-69 & 5.49 & 0.34 & 0.07 & 0.97\% \\
        70+ & 5.56 & 0.34 & 0.06 & 0.92\% \\
        \bottomrule
    \end{tabular}
    \caption{Temporal mean and standard deviation of the weekly \emph{individual compliance score} time series, by age group. Also shown: Reconstruction root mean squared (RMSE) and mean average percentage error (MAPE) of the reconstruction of the respective time series based on the universal (collapsed) time series $y_t$.}
    \label{tab:reconstruction-individual-behaviour}
\end{table}

\begin{table}
    \centering
    \begin{tabular}{lrrrr}
        \toprule
         & \textbf{Mean} & \textbf{Std. Dev.} & \textbf{RMSE} & \textbf{MAPE}\\
        \midrule
        18-29 & 5.05 & 0.15 & 0.06 & 0.88\% \\
        30-39 & 5.17 & 0.14 & 0.05 & 0.83\% \\
        40-49 & 5.28 & 0.12 & 0.05 & 0.66\% \\
        50-59 & 5.46 & 0.10 & 0.05 & 0.81\% \\
        60-69 & 5.70 & 0.08 & 0.07 & 0.90\% \\
        70+ & 5.77 & 0.08 & 0.06 & 0.75\% \\
        \bottomrule
    \end{tabular}
    \caption{Temporal mean and standard deviation of the weekly \emph{authorities trust score} time series, by age group. Also shown: Reconstruction root mean squared (RMSE) and mean average percentage error (MAPE) of the reconstruction of the respective time series based on the universal (collapsed) time series $y_t$.}
    \label{tab:reconstruction-authorities-trust}
\end{table}

\begin{table}
    \centering
    \begin{tabular}{lrrrr}
        \toprule
         & \textbf{Mean} & \textbf{Std. Dev.} & \textbf{RMSE} & \textbf{MAPE}\\
        \midrule
        18-29 & 3.90 & 0.41 & 0.10 & 2.29\% \\
        30-39 & 3.83 & 0.42 & 0.10 & 2.37\% \\
        40-49 & 3.76 & 0.42 & 0.07 & 1.53\% \\
        50-59 & 3.67 & 0.44 & 0.06 & 1.29\% \\
        60-69 & 3.64 & 0.44 & 0.09 & 1.99\% \\
        70+ & 3.88 & 0.44 & 0.14 & 3.03\% \\
        \bottomrule
    \end{tabular}
    \caption{Temporal mean and standard deviation of the weekly \emph{life degradation score} time series, by age group. Also shown: Reconstruction root mean squared (RMSE) and mean average percentage error (MAPE) of the reconstruction of the respective time series based on the universal (collapsed) time series $y_t$.}
    \label{tab:reconstruction-life-degradation}
\end{table}

\begin{figure}
    \centering
    \includegraphics[width=\textwidth]{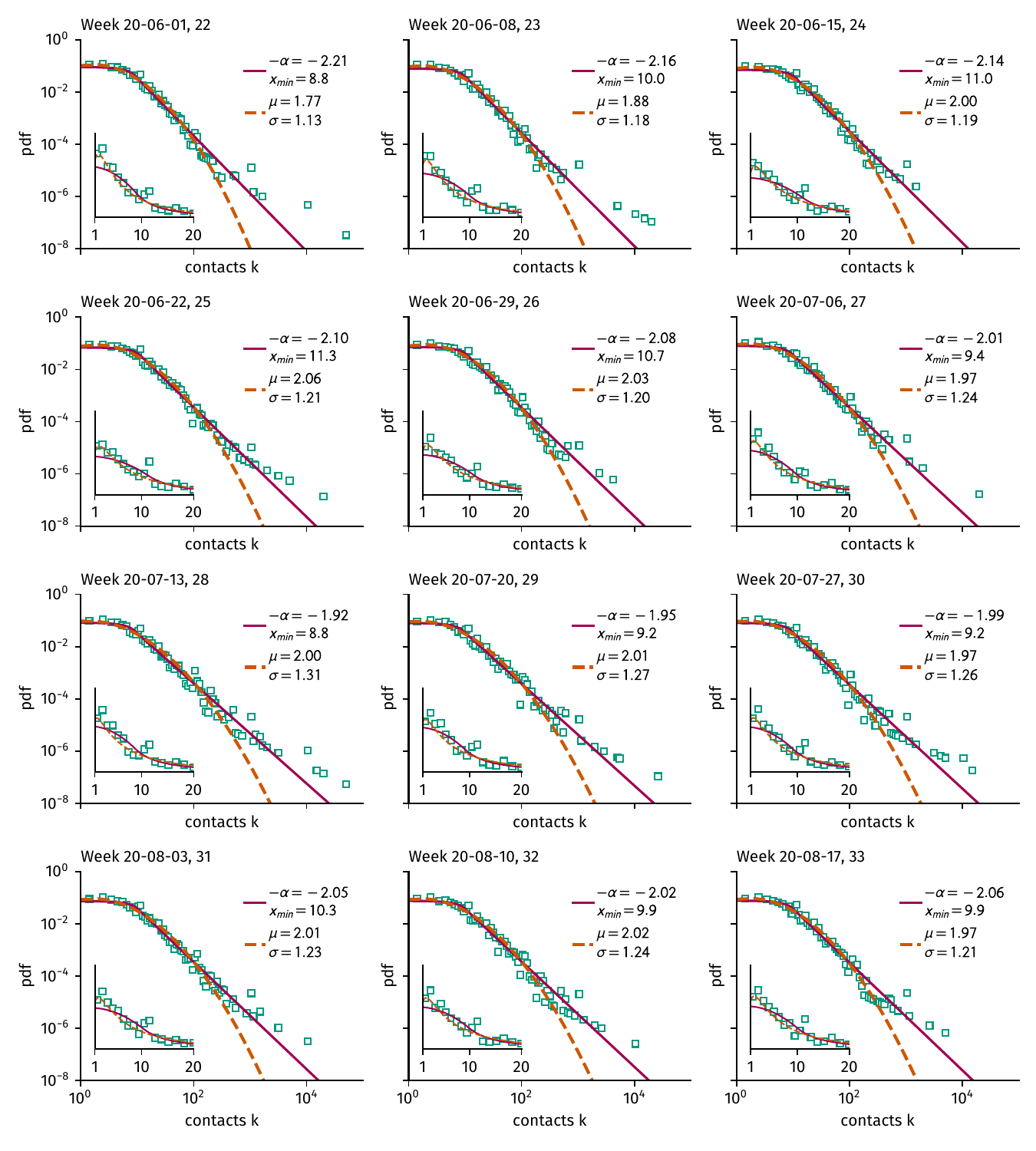}
    \caption{Contact distributions for values pooled over calendar weeks, neglecting reported values $k\leq0$ and $k\geq10^6$. We show fits of a log-normal distribution (dashed orange line) and Pareto distribution with algebraic core (magenta). Insets show the core of the distribution.}
    \label{fig:weekly-contact-pdf-00}
\end{figure}

\begin{figure}
    \centering
    \includegraphics[width=\textwidth]{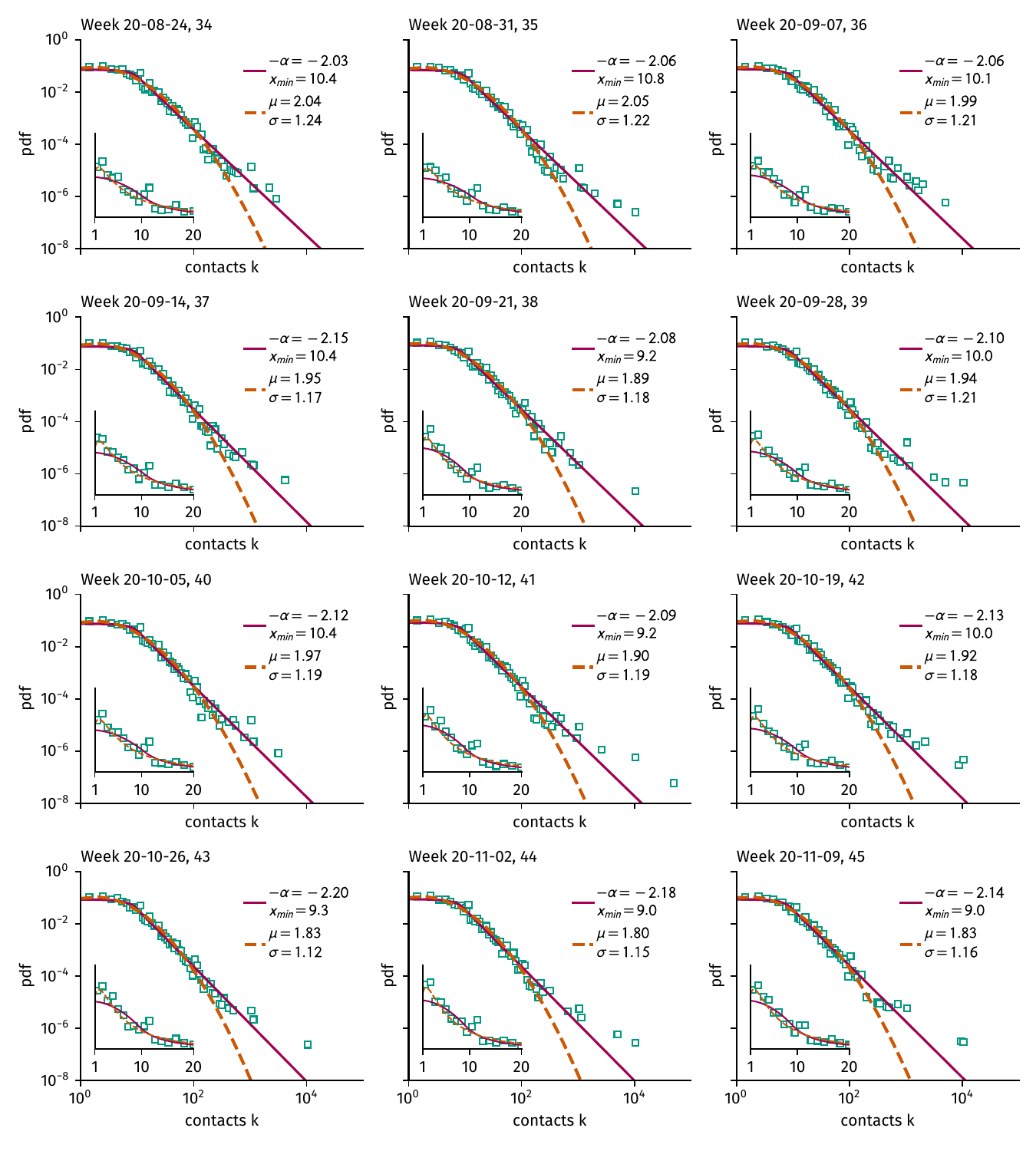}
    \caption{Weekly contact distributions with fits, continuation.}
    \label{fig:weekly-contact-pdf-01}
\end{figure}

\begin{figure}
    \centering
    \includegraphics[width=\textwidth]{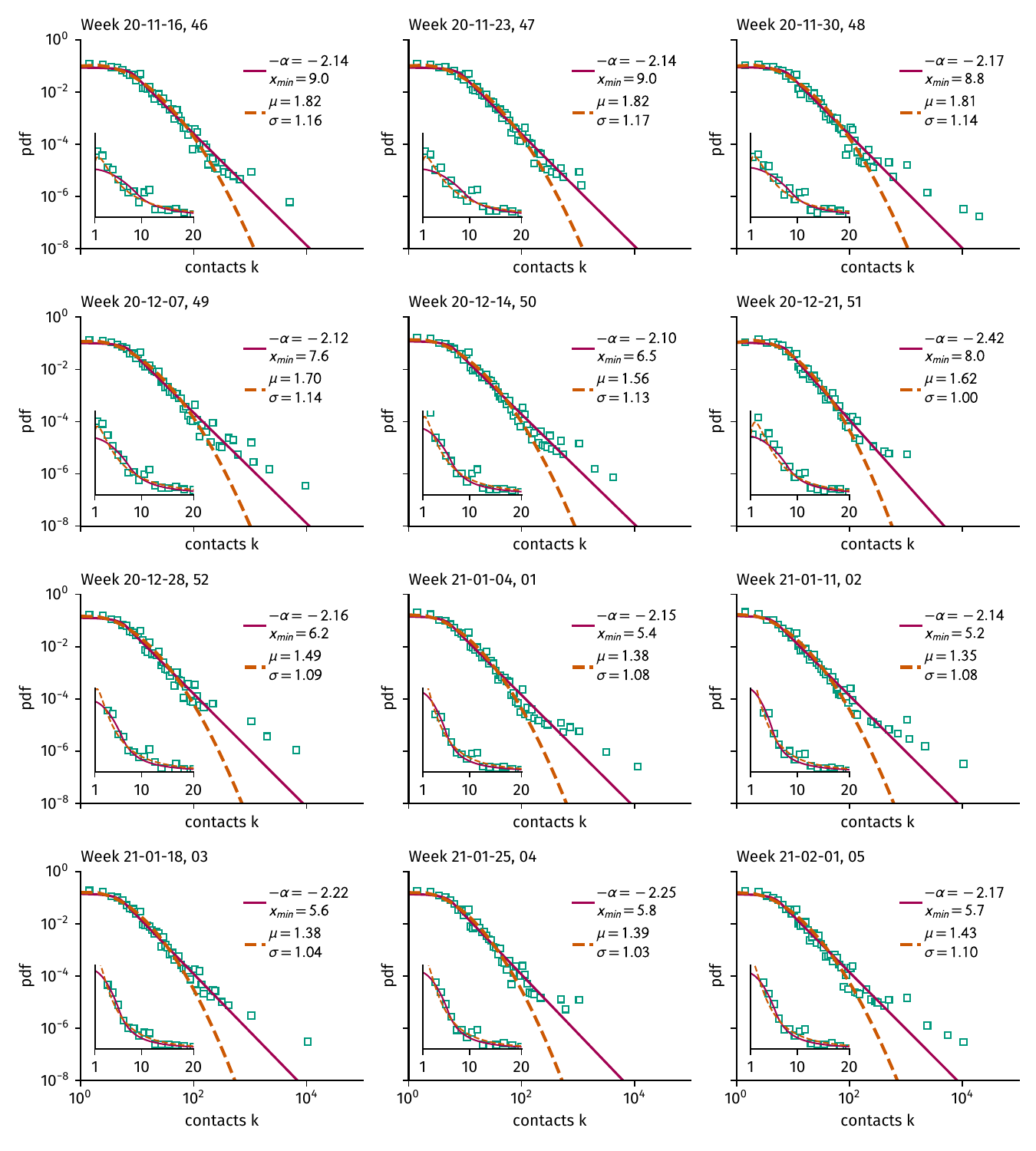}
    \caption{Weekly contact distributions with fits, continuation.}
    \label{fig:weekly-contact-pdf-02}
\end{figure}

\begin{figure}
    \centering
    \includegraphics[width=\textwidth]{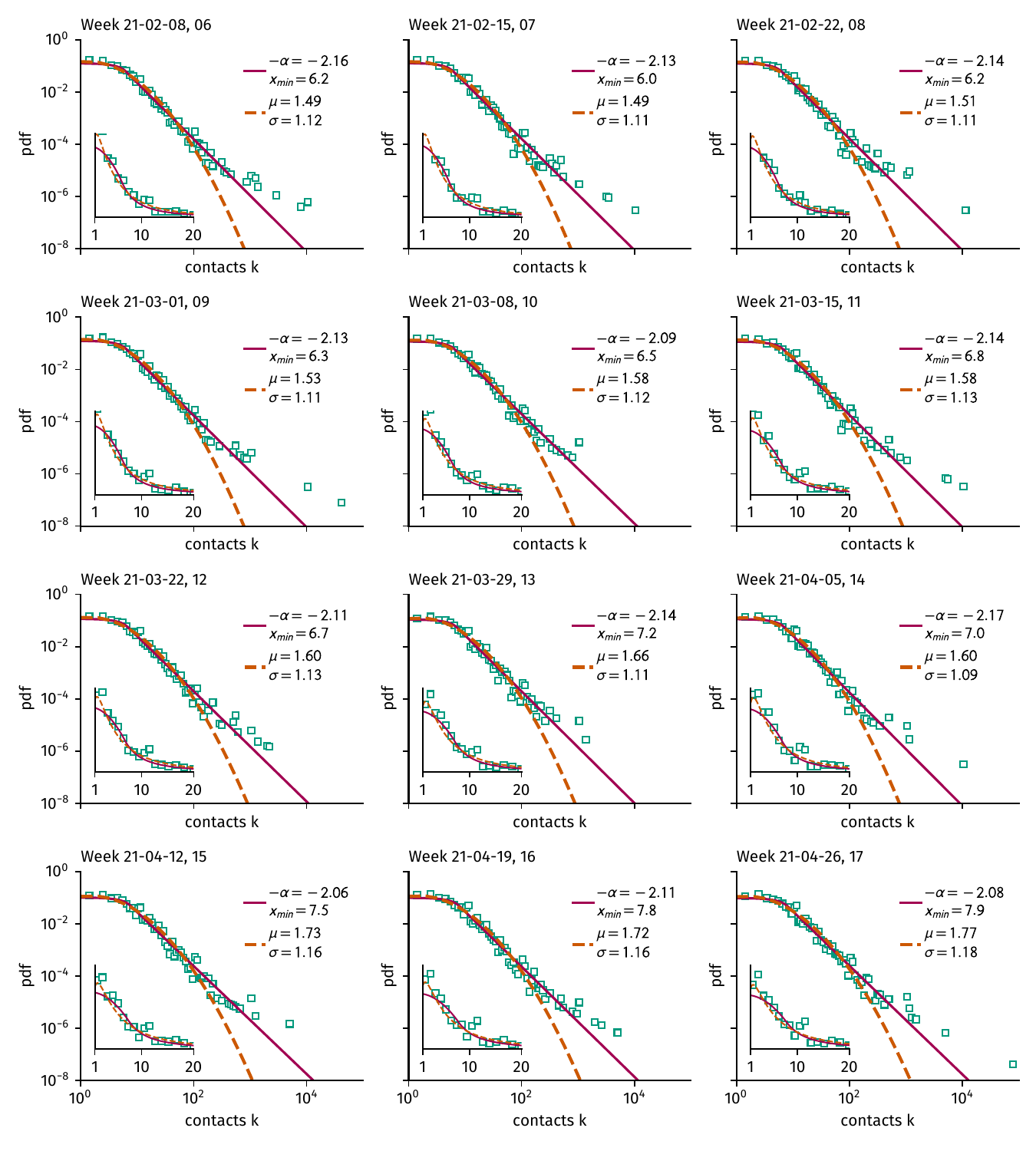}
    \caption{Weekly contact distributions with fits, continuation.}
    \label{fig:weekly-contact-pdf-03}
\end{figure}

\begin{figure}
    \centering
    \includegraphics[width=\textwidth]{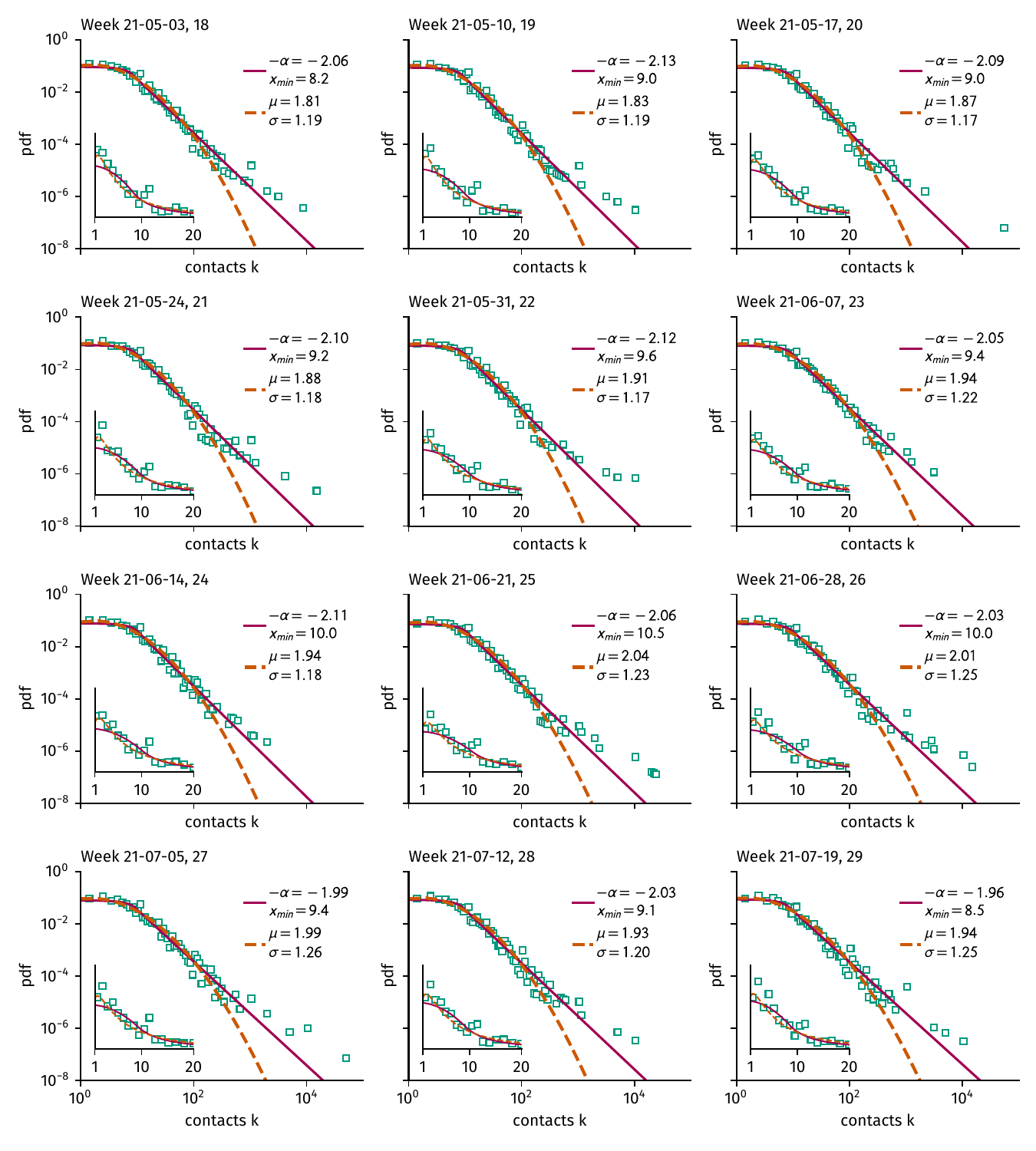}
    \caption{Weekly contact distributions with fits, continuation.}
    \label{fig:weekly-contact-pdf-04}
\end{figure}

\begin{figure}
    \centering
    \includegraphics[width=\textwidth]{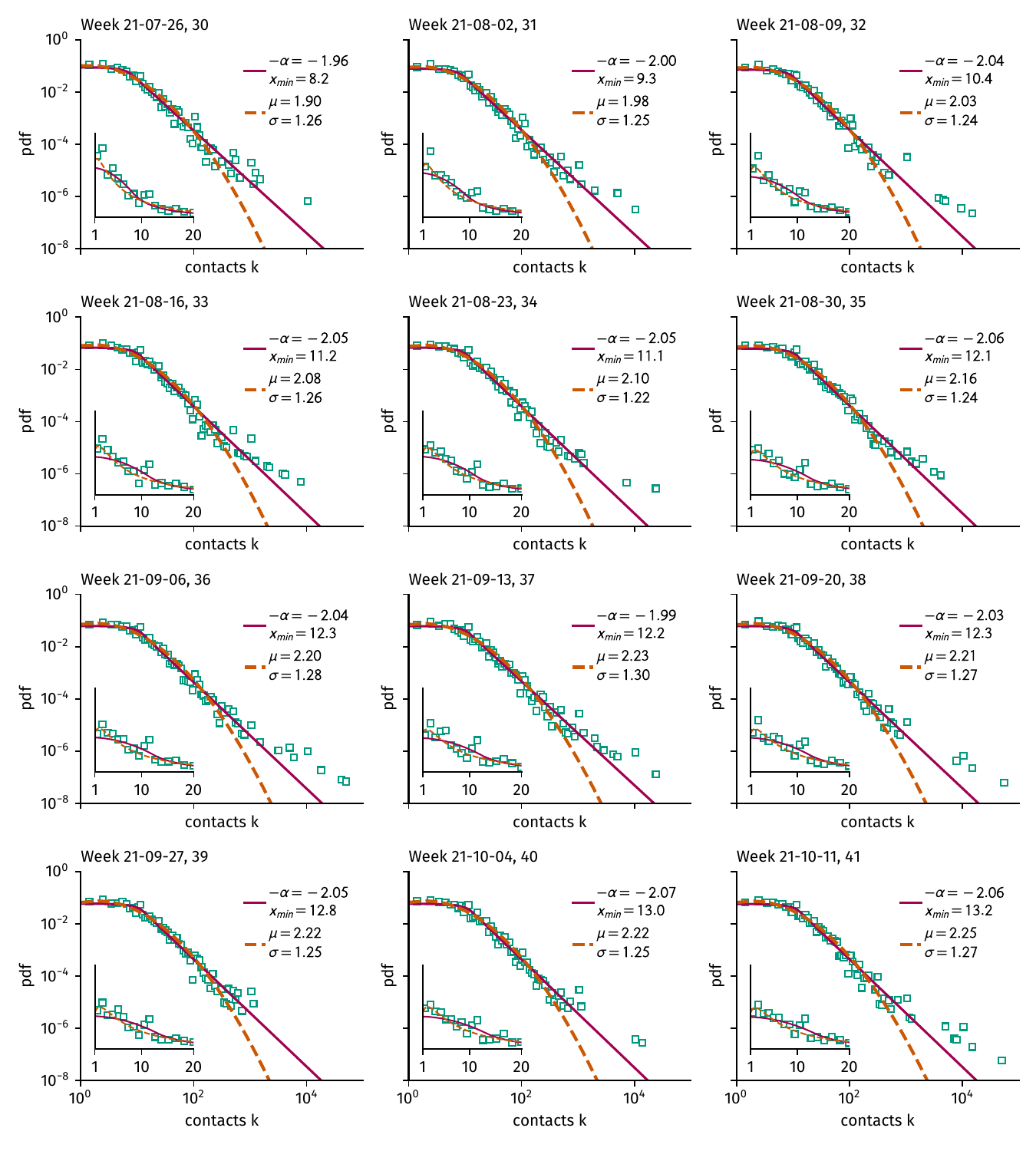}
    \caption{Weekly contact distributions with fits, continuation.}
    \label{fig:weekly-contact-pdf-05}
\end{figure}

\begin{figure}
    \centering
    \includegraphics[width=\textwidth]{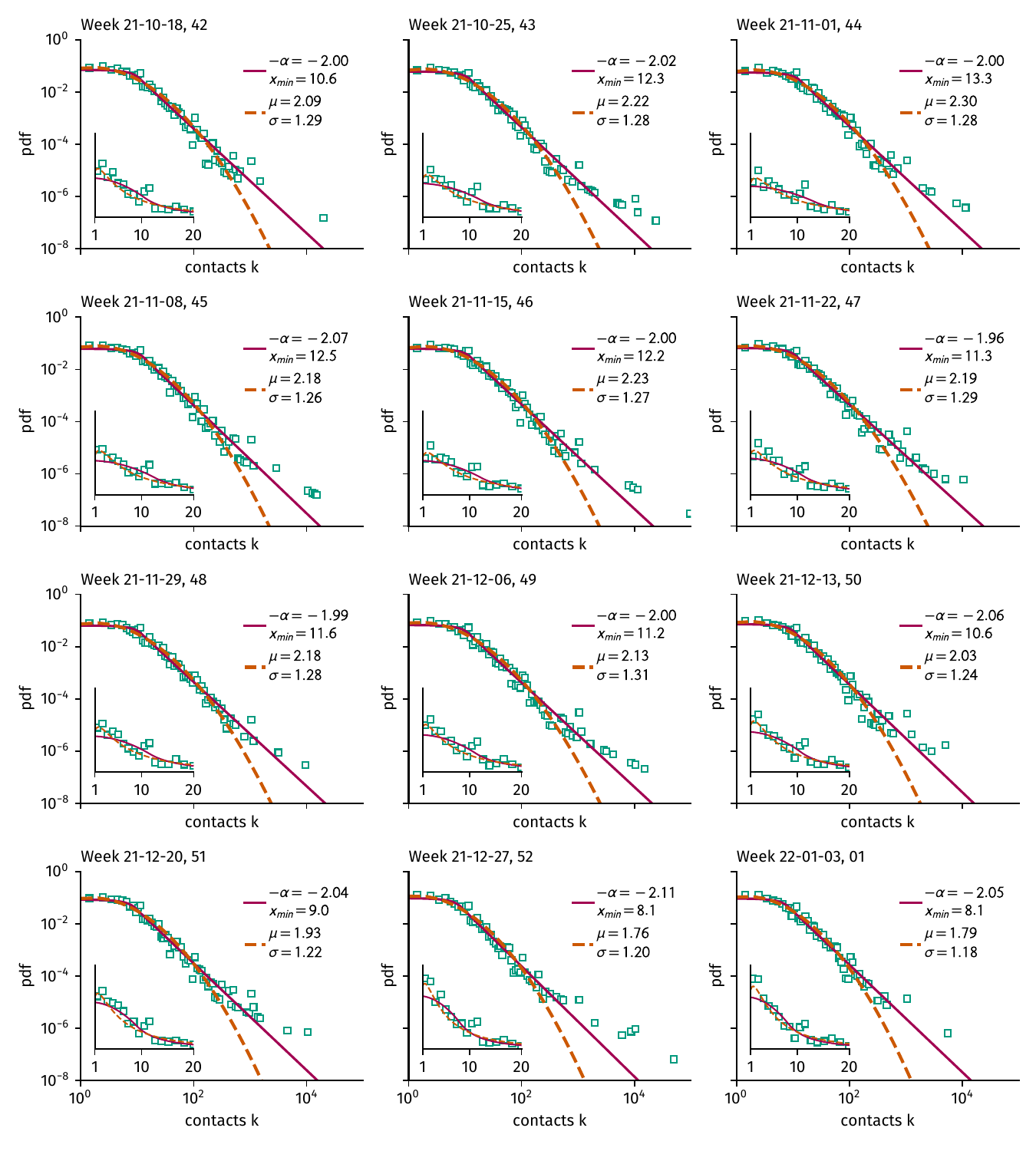}
    \caption{Weekly contact distributions with fits, continuation.}
    \label{fig:weekly-contact-pdf-06}
\end{figure}

\begin{figure}
    \centering
    \includegraphics[width=\textwidth]{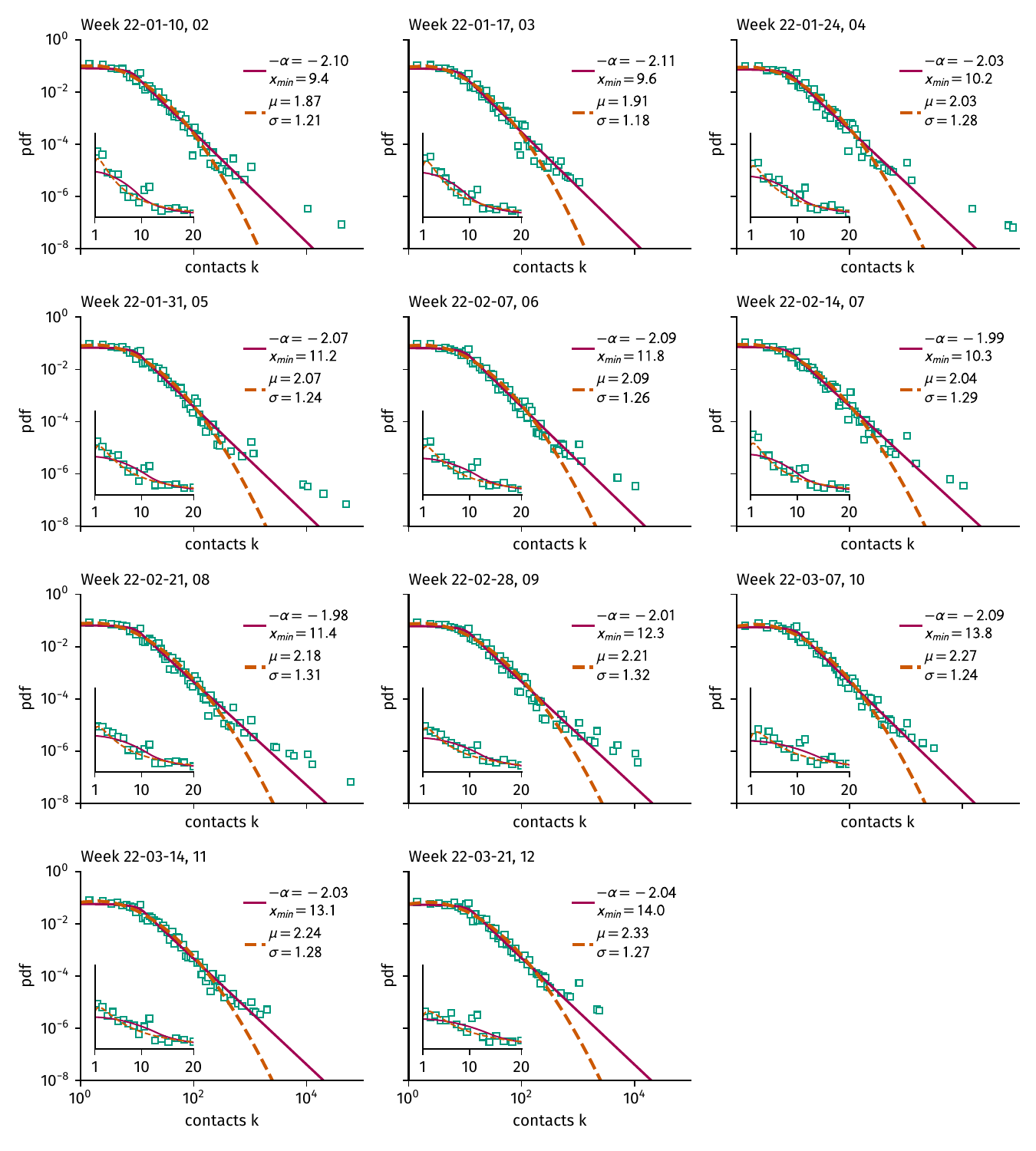}
    \caption{Weekly contact distributions with fits, continuation.}
    \label{fig:weekly-contact-pdf-07}
\end{figure}

\begin{figure}
    \centering
    \includegraphics[width=0.45\textwidth]{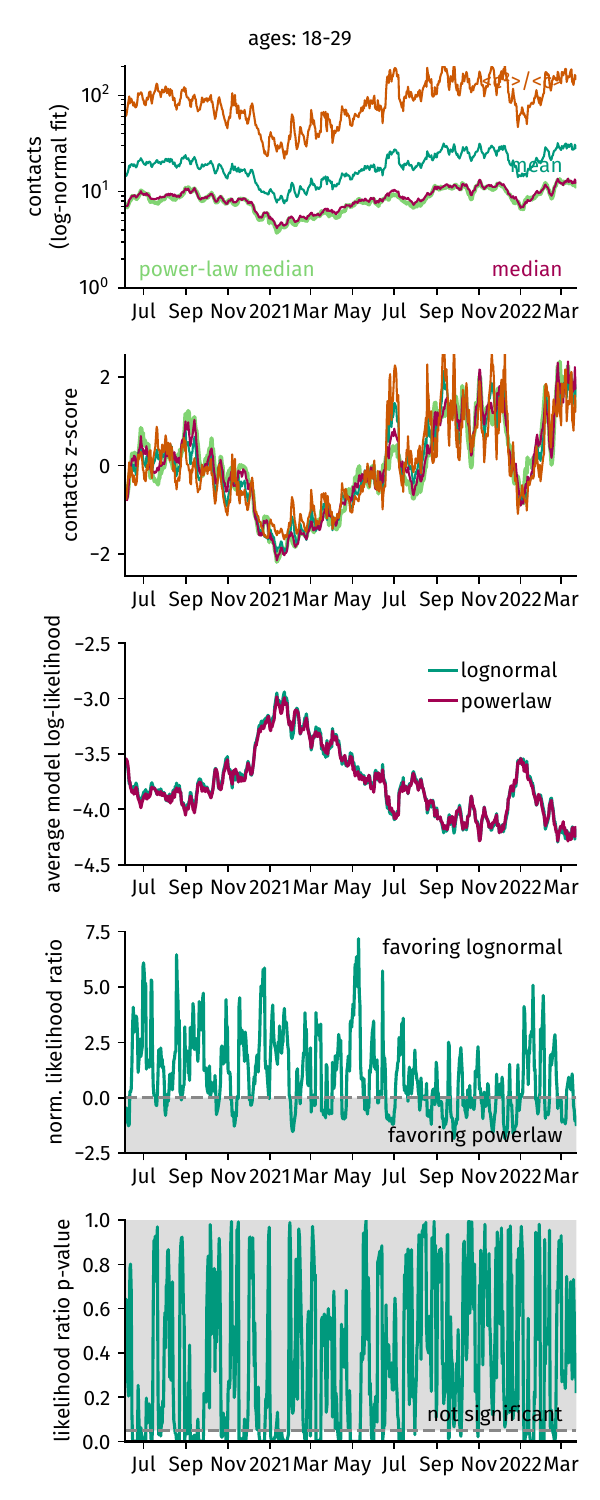}
    \includegraphics[width=0.45\textwidth]{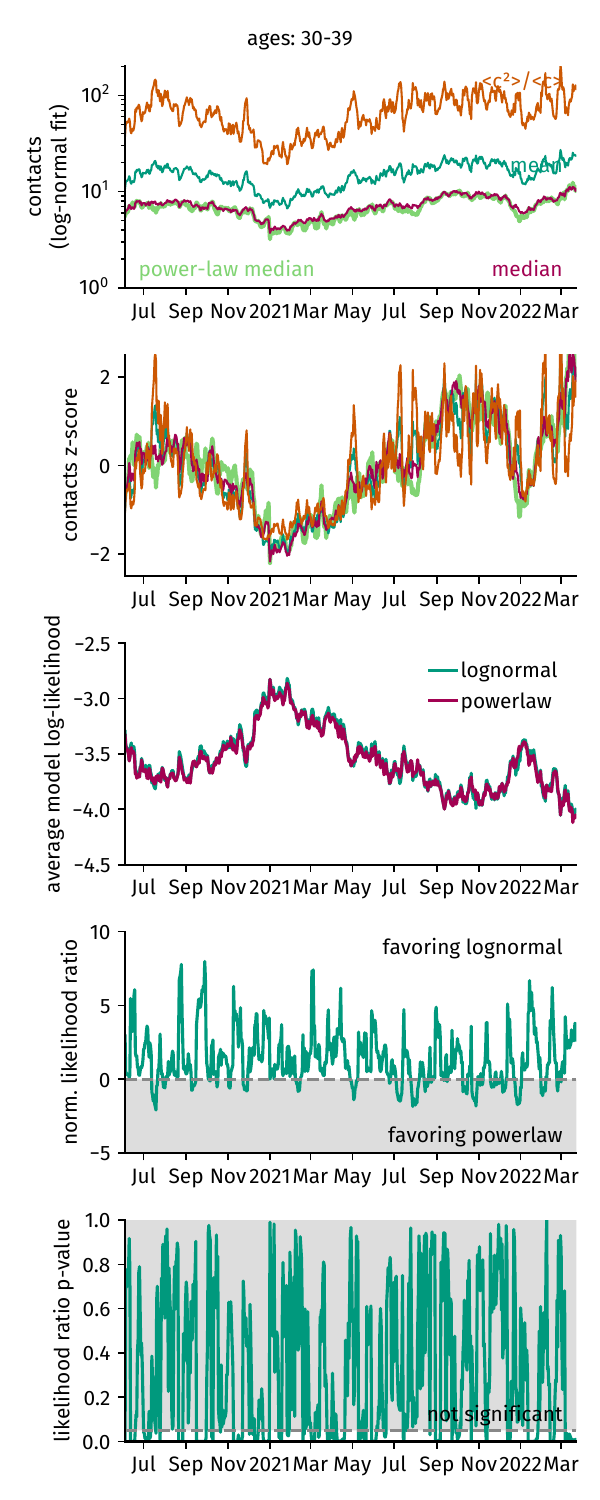}
    \caption{Results of fitting a log-normal and a Pareto distribution with algebraic core to data pooled over a symmetric 7-day sliding window around day $t$. We see that mostly, the log-normal distribution is favored over the Pareto distribution, with exceptions usually being statistically insignificant. The median approximately follows the same form for both distributions. The temporal evolution of all contact observables is sufficiently similar. Shown here are age groups 18--29 and 30-39.}
    \label{fig:daily-age-group-pdfs-18-39}
\end{figure}

\begin{figure}
    \centering
    \includegraphics[width=0.45\textwidth]{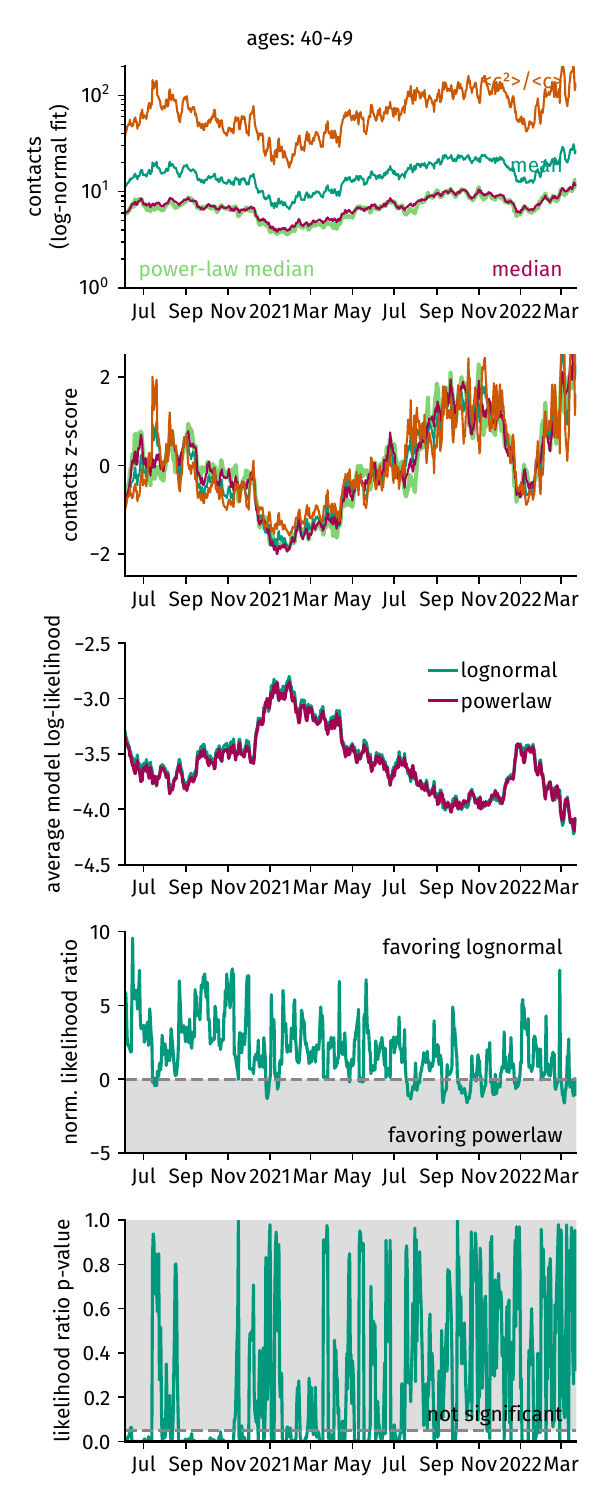}
    \includegraphics[width=0.45\textwidth]{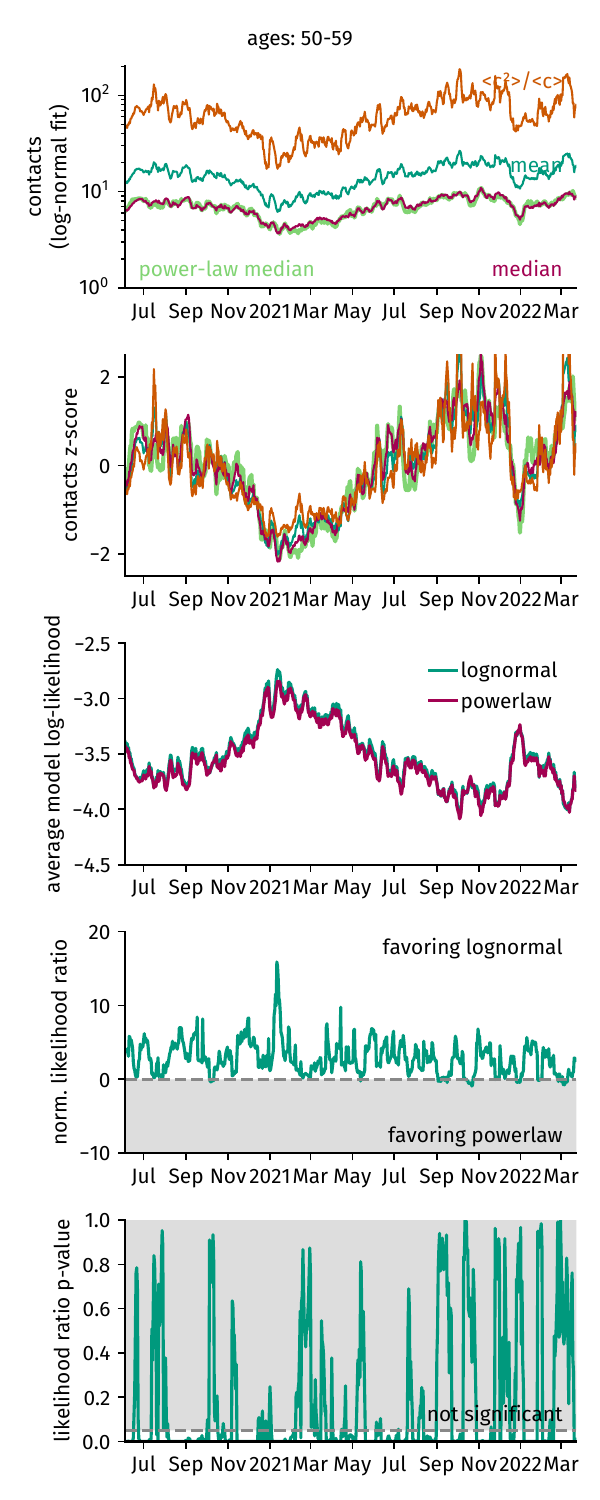}
    \caption{Temporal evolution of contact distribution fits for age groups 40--49 and 50--59.}
    \label{fig:daily-age-group-pdfs-40-59}
\end{figure}

\begin{figure}
    \centering
    \includegraphics[width=0.45\textwidth]{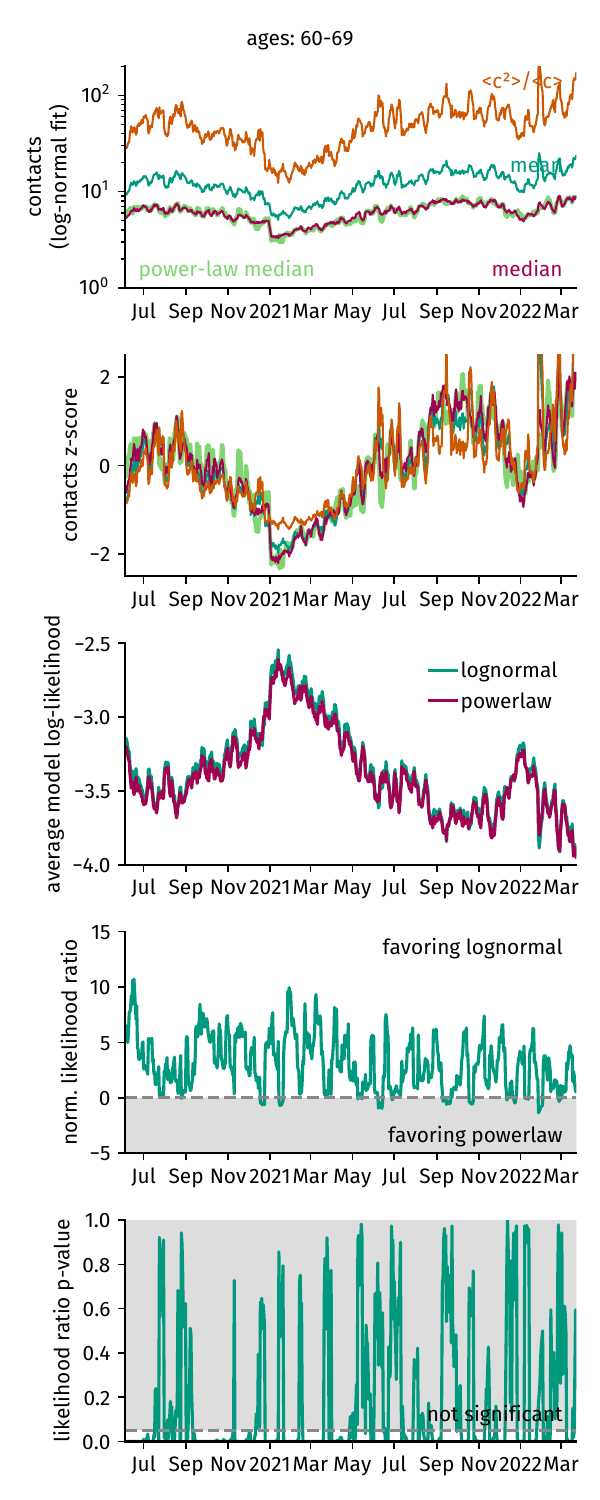}
    \includegraphics[width=0.45\textwidth]{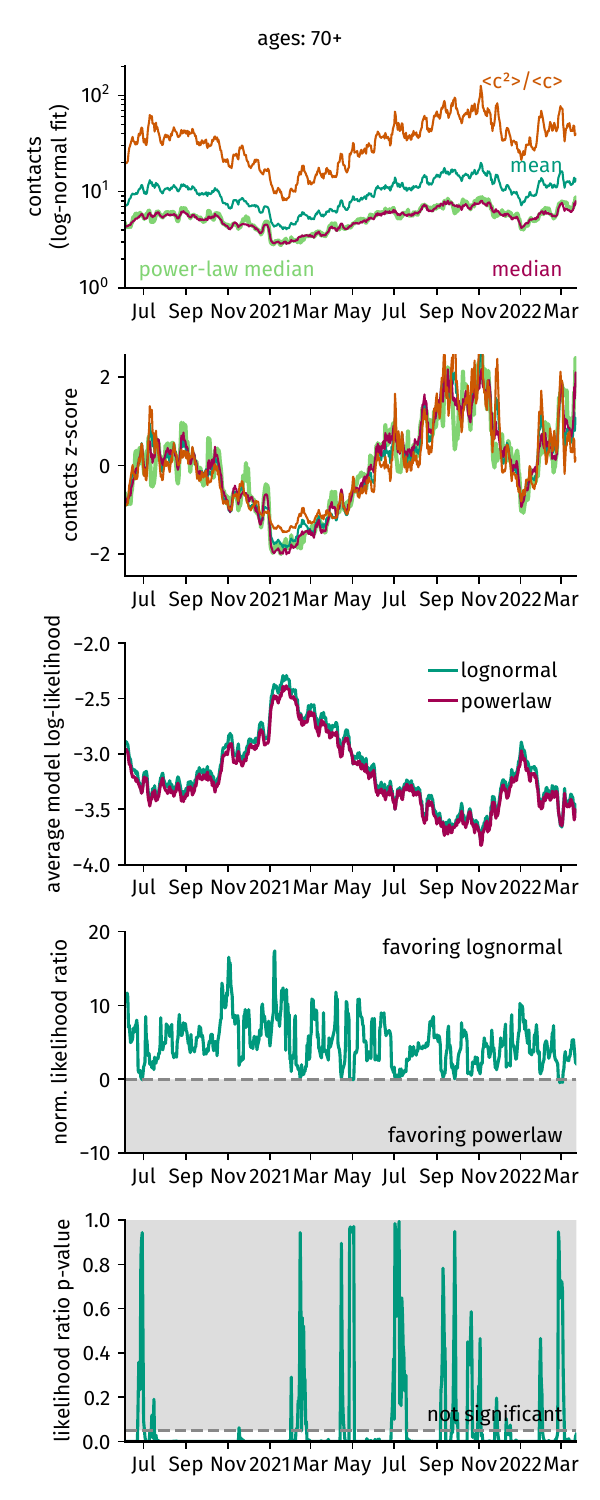}
    \caption{Temporal evolution of contact distribution fits for age groups 60--69 and 70+.}
    \label{fig:daily-age-group-pdfs-60+}
\end{figure}

\begin{figure}
    \centering
    \includegraphics[width=0.45\textwidth]{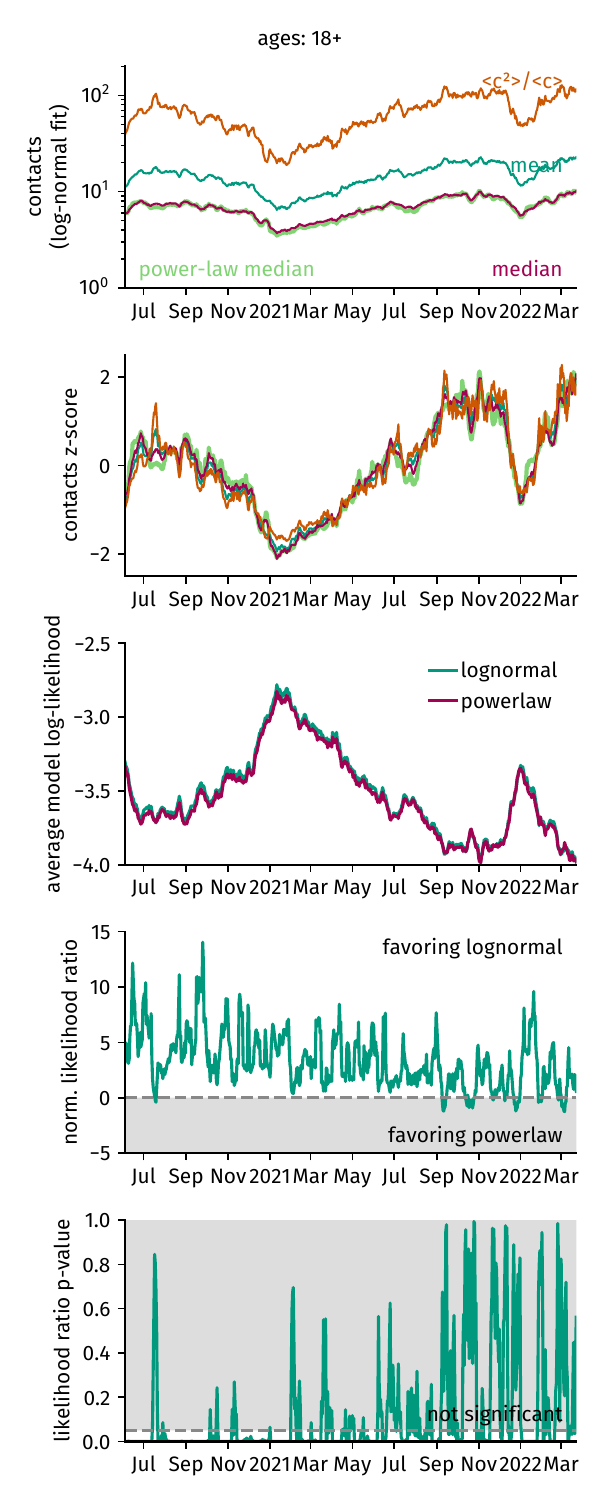}
    \caption{Temporal evolution of contact distribution fits for all ages.}
    \label{fig:daily-age-group-pdfs-18+}
\end{figure}

\begin{figure}
    \centering
    \includegraphics[width=0.85\textwidth]{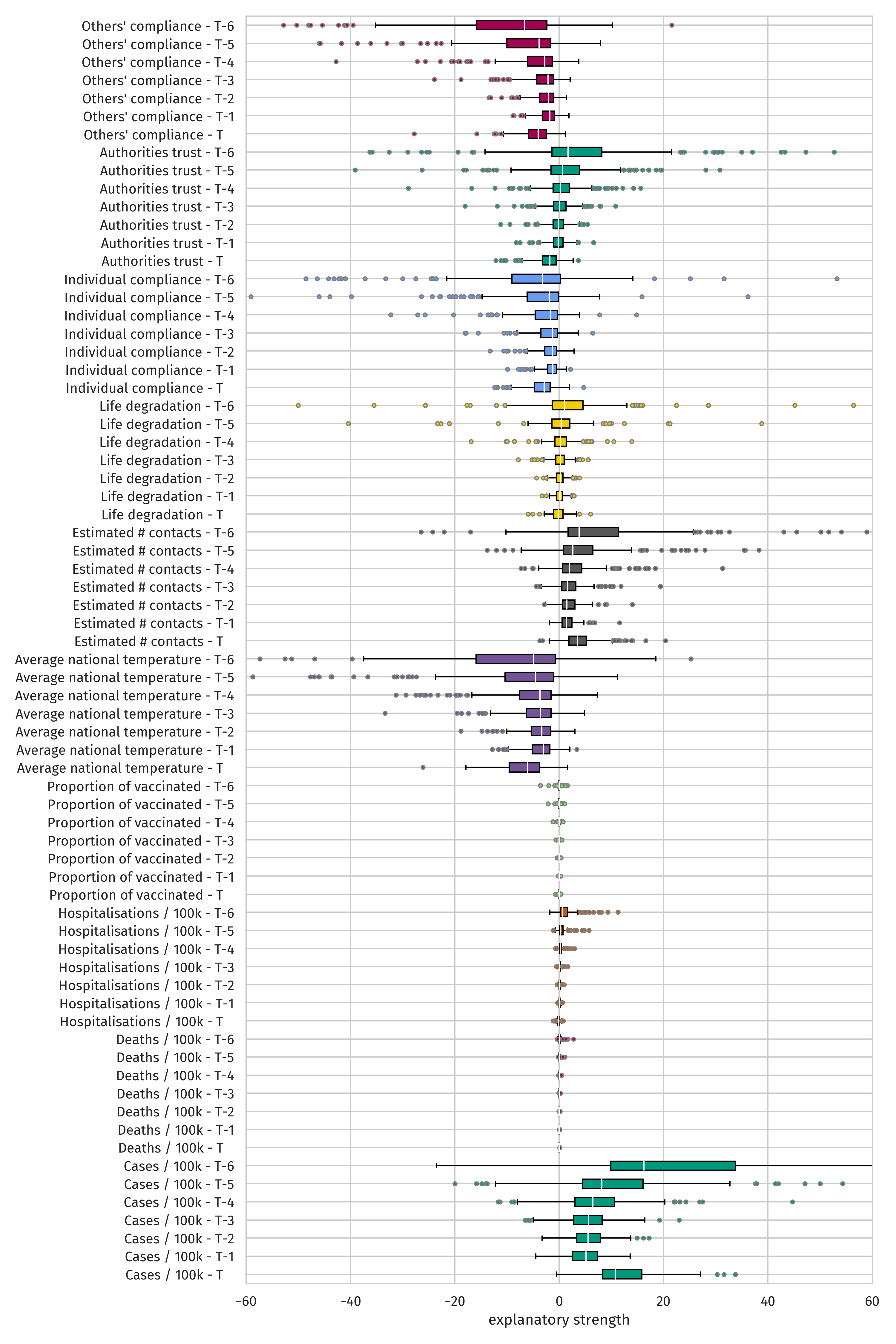}
    \caption{Relevance score distribution for all input features in the full LSTM model, depending on the day of the input. Each relevance score value represents the relevance of the respective input feature on day $t-\tau$ for the prediction on day $t+14$, averaged over the whole time series. Distributions are amassed from 200 independent model fits. We see that all for all features, relevance follows a similar, u-shaped trend of over time. Some outliers lie outside the frame of this figure and are not shown for better visibility of the dominant parts of the distribution.}
    \label{fig:boxplot-explanatory-strength-absolute}
\end{figure}

\begin{figure}
    \centering
    \includegraphics[width=.85\textwidth]{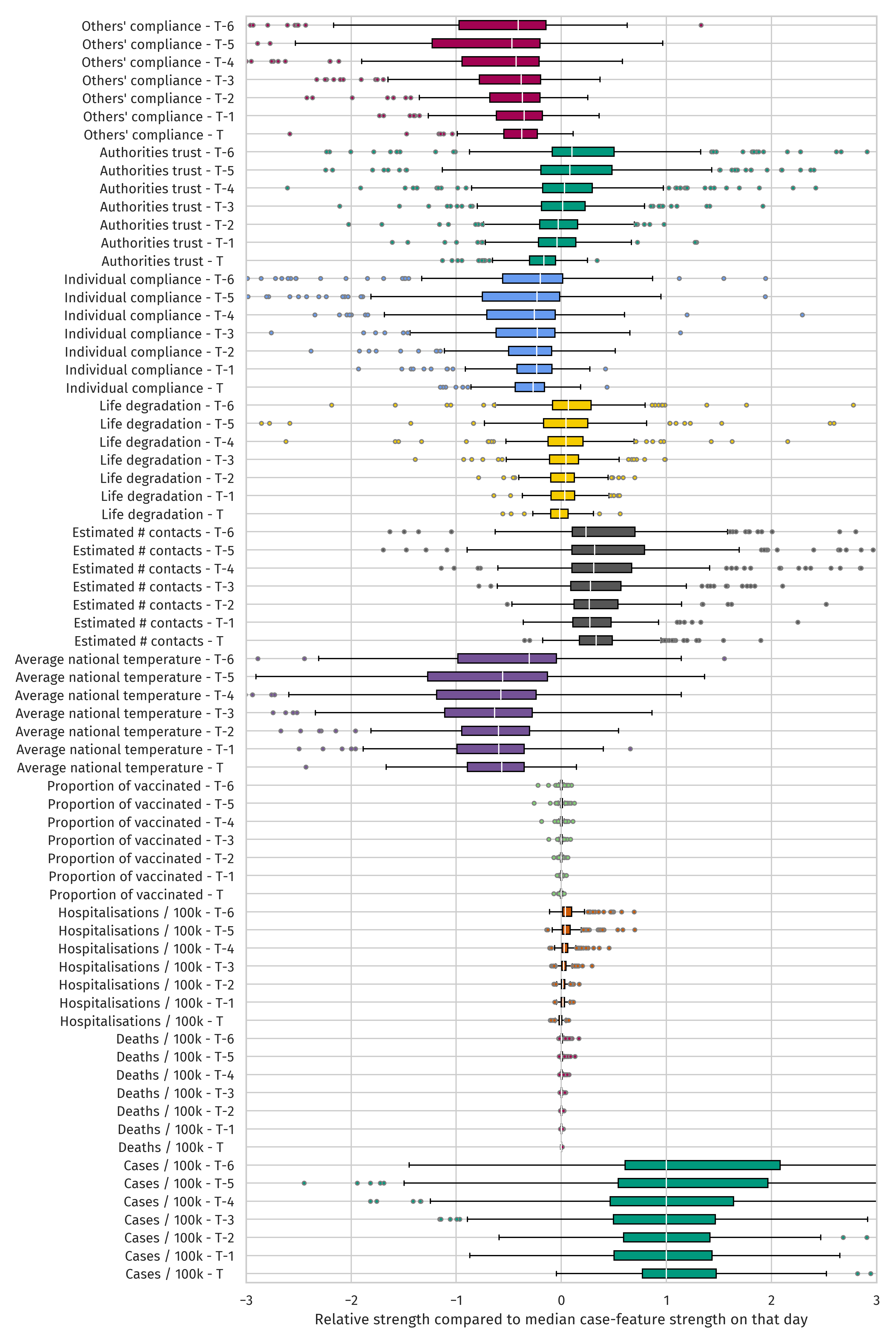}
    \caption{Relative relevance score distribution for all input features in the full LSTM model, depending on the day of the input. The original relevance scores (cf.\ Figure~\ref{fig:boxplot-explanatory-strength-absolute}) have been normalised by the median relevance score of feature ``Cases / 100k'' on the respective day to reduce the u-shape seen in Figure~\ref{fig:boxplot-explanatory-strength-absolute} and thus make distributions more comparable between input days. Some outliers lie outside the frame of this figure and are not shown for better visibility of the dominant parts of the distribution.}
    \label{fig:boxplot-explanatory-strength-relative}
\end{figure}